\documentclass[a4paper]{PoS}
\usepackage[italian,english]{babel}
\usepackage{slashed}
\usepackage{amssymb}
\usepackage{amsfonts}
\usepackage{amsmath}
\usepackage{fancyhdr}
\usepackage{lineno}
\usepackage{subfigure}

\newcommand{\be}{\begin{equation}}
\newcommand{\ee}{\end{equation}}
\newcommand{\beq}{\begin{equation}}
\newcommand{\eeq}{\end{equation}}
\newcommand{\beqa}{\begin{eqnarray}}
\newcommand{\eeqa}{\end{eqnarray}}

\newcommand{\bea}{\begin{eqnarray}}
\newcommand{\eea}{\end{eqnarray}}

\newcommand{\lsim}{\mathrel{\mathop{\kern 0pt \rlap
  {\raise.2ex\hbox{$<$}}}
  \lower.9ex\hbox{\kern-.190em $\sim$}}}
\newcommand{\gsim}{\mathrel{\mathop{\kern 0pt \rlap
  {\raise.2ex\hbox{$>$}}}
  \lower.9ex\hbox{\kern-.190em $\sim$}}}

\usepackage{pstricks}
\usepackage{color}
\usepackage{multirow}
\usepackage{cite}
\usepackage{braket}

      \let\g=\gamma

\title{ Dark Matter with Light and Ultralight St\"uckelberg Axions}

\ShortTitle{Axion dark matter }

\author{\speaker{Claudio Corian\`o}\\
INFN Sezione di Lecce, 
        Dipartimento di Matematica e Fisica, \\ Universit\`a del Salento, \\ Via Arnesano, 73100 Lecce, Italy\\
E-mail: \email{claudio.coriano@le.infn.it}}

\author{ Matteo Maria Maglio, Alessandro Tatullo, Dimosthenis Theofilopoulos\\
   INFN Sezione di Lecce, 
        Dipartimento di Matematica e Fisica, \\ Universit\`a del Salento, \\ Via Arnesano, 73100 Lecce, Italy\\
        E-mail: \email{matteomaria.maglio@le.infn.it, alessandro.tatullo@le.infn.it,dimosthenis.theofilopoulos@le.infn.it \\}}

\abstract{Scenarios with axion-like particles of variable masses, from the milli-eV range to ultralight, can be generated in a natural way in the context of effective field theory models with anomalous $U(1)$ gauge symmetries. They include dimension-5 operators which define a coupling of the axion to the gauge anomaly. They provide a realization, in the domain of ordinary gauge theories, of models for such particles, which evade the usual mass/coupling constraint of ordinary 
(Peccei-Quinn) axions and are natural dark matter candidates. As an illustration of these models, we present an overview of two of these scenarios. One of them is built around the next-to-minimal MSSM (NMSSM), a model called the
USSM-A, which illustrates how the St\"uckelberg supermultiplet can be used to generate two dark matter candidates, a neutralino containing an axino component coming from the St\"uckelberg sector, plus the axion ($\textrm{Im}\, b$). The real component of the complex St\"uckelberg 
field carries dilaton-like ($\textrm{Re}\, b\, FF$) interactions. In a second model, 
non-supersymmetric, the St\"uckelberg scale is raised up to the GUT epoch. In this case the axion mass can be ultralight ($\sim 10^{-20}$ eV). The periodic potential generated at the GUT phase transition and the corresponding oscillations are related to a particle whose De Broglie wavelength can be sub-galactic. A similar analysis is also possible for the supersymmetric scenario. }

\FullConference{Proceedings of the Corfu Summer Institute 2019 "School and Workshops on Elementary Particle Physics and Gravity"\\

                 1-27 September 2019\\

                 Corfu, Greece}

\begin{document}
\section{Introduction}
Astrophysical and cosmological data have been providing dramatic evidence that about $\sim 80 \%$ of matter/energy in the universe is in an unknown form, presenting a major challenge for both 
theoretical and experimental searches. In particular, approximately a quarter of such missing content is supposed to be accounted for by dark matter. \\
Candidates for dark matter abound, from 
heavy and diluted states, in the form of relics produced in the early universe, to light or even ultralight particles which also have decoupled at an early stage of the evolution of the cosmo. 
Analysis of the velocities of stars orbiting galaxies, from their rotation curves, or precision studies of the cosmic microwave background, indicate that progress in this field has been remarkably 
steady with an overwhelming amount of information collected both by terrestrial and satellite experiments. 
Such a large amount of data have been confirming with significant accuracy the 
 standard $\Lambda$CDM dark matter/dark energy cosmological model \cite{Ade:2015xua}, which has been very successfull in explaining most of these results. \\
 According to $\Lambda$CDM, the dark energy component provides about $68\%$ of the total mass/density contributions of our universe and can take the form of a cosmological constant. Whether this corresponds only to a successfull phenomenological description of the dark energy or other, remains an open issue. Dark energy dominance in the cosmological expansion at late times can account for the cosmological acceleration measured by Type Ia supernovae \cite{Riess:1998cb,Perlmutter:1998np}, with ordinary baryonic dark matter contributing just a few percent of the total mass/energy content ($\sim 5 \%$). A smaller neutrino component can also be part of 
the proposed distributions. \\
Challenging as they are, the answers to such questions force us to take into consideration all the possible information that can be deduced from particle model building, following primarily the idea of unification. \\
 In this work we are going to overview one candidate for dark matter which is directly 
 associated with anomaly actions, under the assumption that an anomalous abelian symmetry could provide ground for its existence. As discussed in several older and also in more recent works, the construction that we outline defines a rather simple framework for the generalization of axion models. At a field theory level, our models are based on the assumption that there is  an underlying anomalous abelian gauge symmetry in the theory, in which a St\"uckelberg pseudoscalar $(b(x))$ restores the gauge invariance of the Lagrangian, broken at 1-loop by the interactions of the anomalous $U(1)$ current.\\
Like any ordinary Goldstone mode, the St\"uckelberg will undergo a shift under the $U(1)$ transformations, which will be local, for being associated with a gauge symmetry. 
The breaking of such shift symmetry allows to identify one physical component of the St\"uckelberg field. This is the result of the combined action of the symmetry breaking mechanism from the scalar sector, at a specific scale, coming from the ordinary (polynomial) part of the potential, and of an extra (non perturbative) periodic potential, as in the Peccei-Quinn (PQ) case.\\
The breaking will induce a mixing between the Higgs and the St\"uckelberg, which appears as a phase in a nonperturbative (extra) potential, generated at a specific phase transition and assumed to be of instanton origin. Indeed, the generation of such periodic potential is generic in non abelian gauge theories, and it is expected to be exponentially suppressed, with a suppression $(\lambda)$ which is controlled by the value of the running coupling at the transition scale $Q_T$ $(\lambda \sim e^{-2\pi/\alpha(Q_T)})$.

\section{A large parameter space for masses and couplings}
Models of this type allow to evade the constraints between the axion mass and the coupling of the axion to the anomaly, which severely limits the parameter space of an ordinary 
PQ axion, whose properties are controlled by a single constant $(f_a)$. This significant extension of the parameter space is due to the different mechanism introduced for the cancellation of the anomaly, which requires a Goldstone mode associated with a gauge rather than to a global symmetry. The Wess-Zumino interactions appearing in the extended Lagrangian, which allow to qualify the pseudoscalar as an axion, appear as counterterms for the restoration of the gauge invariance of the theory, in a local, non-renormalizable action with a typical anomaly term.
\subsection{Anomalies and anomaly actions}
In the SM, anomaly cancellation is enforced by charge assignment, by a direct balance between the left and right chirality modes of the fermion spectrum. In the models that we address, instead, this balance is lost and the gauge invariance of the anomaly effective action is a result of the exchange of a physical pseudoscalar. In general, an anomaly action embodies the variation of a certain symmetry by the inclusion of a certain Goldstone mode (scalar/pseudoscalar) which couples to a global anomaly. \\
For instance, {\em local} conformal anomaly actions require one extra degree of freedom in the spectrum, a dilaton, and reproduce a certain anomaly functional by the variation of the dilaton field under the anomalous symmetry. Here the adjective "local" refers to the fact that there are no terms such as $1/\square$ or Green's functions in general, in their defining expressions. \\
The coupling of the dilaton manifests with up to quartic interactions \cite{Coriano:2012dg} in the local versions of such actions. Their nonlocal versions, instead - in the conformal case- introduce the Green functions of quartic conformally covariant operators \cite{Coriano:2017mux}. In the case of a gauged anomalous $U(1)$ current, the restoration of gauge symmetry, which in this case is necessary, can be obtained just with a linear coupling of a pseudoscalar field to the anomaly.  It is clear, though, that anomaly actions are not uniquely defined, and can be characterized by different completions at high energy. \\
For example, if we start from an anomaly-free fermion multiplet and decouple one of the chiral fermions, for instance by rendering it heavy via a large Yukawa operator, the effective Lagrangian that we are left with at low energy defines an anomaly action with interactions which are of Wess-Zumino type. This observation has has been exploited recently in \cite{Coriano:2019vjl}, in the context of the decoupling of a right-handed neutrino in a $SO(10)\times U(1)_X$ scenario, with the possibility of generating a pseudoscalar particle of GUT 
origin as a possible dark matter candidate. \\   
For these reasons, St\"uckelberg Lagrangians can be the byproduct at low energy both of string theory completions - for instance as possible realization of a mechanism of anomaly cancellation, with the St\"uckelberg being the dual of a 3-form - or, more simply, from the decoupling of a chiral fermion in an (anomaly-free) irreducible multiplet of chiral fermions.  \\  
As in the conformal case, also in the chiral case the effective action can be nonlocal, and requires a suitable completion in order to be consistently defined.  
This point has been previously reviewed by us in \cite{Coriano:2019dyc}. 

Considering the chiral case, in the low energy theory there are surviving trilinear gauge interactions which are specific of this class of actions, involving an anomalous extra Z prime, whose mass is given by the St\"uckelberg scale.
The inclusion of such extra anomalous symmetry close to the Planck or GUT (Grand Unification Theory) scales, is what 
allows to generate an ultralight axion, which can be of the order of $10^{-20}$ eV, 
while the anomalous gauge boson decouples. \\
In fact, in general, in these models the mass of the axion is given by the product of two independent scales, $M_{\textrm {GUT}}$, the GUT scale, and the size of the periodic instanton potential in the GUT theory $(\lambda)$, which is strongly exponentially suppressed and generated at the GUT phase transition.\\
The fact that the mass of the ultralight axion is comparable with the value required in order to solve various issues at subgalactic scale in their matter distributions, provides support for such scenario.
In the next few sections we will be discussing at first a non-supersymmetric version of such models, turning to their supersymmetric extensions in the second part of our review. We will conclude our overview with an illustration of more recent developments in this area, with the formulation of a model around the GUT scale, with a $E_6\times U(1)_X$ symmetry, which predicts two axions: one of them is an ordinary PQ axion, while the second is of St\"uckelberg origin and it is ultralight. \\    
The constructions that we review are all characterised by similar operatorial contents, independently of the scale at which they are supposed to be defined, and should be viewed as a class 
of gauge invariant local effective actions which describe scenarios with 
a different mechanism of anomaly cancellation at work. Axion-like particle find their simplest and well-defined appearance in this class of models, even if their defining Lagrangians are characterised by very different high energy completions.

 \section{Axions}
PQ axions have been studied for several decades as a realistic attempt to solve the strong CP problem
\cite{Peccei:1977ur,Peccei:2006as},\cite{Weinberg:1977ma,Wilczek:1977pj,Dine:1981rt,Zhitnitsky:1980tq,Kim:1979if}\cite{Shifman:1979if}, with the possibility of accounting for the missing dark matter. Ultralight axions, which can be considered their variants, have been originally suggested also as a possible solution of the dark energy problem for an axion mass of $\sim 10^{-33 }$ eV and smaller, which could be generated at the electroweak phase transition \cite{Nomura:2000yk}. In this case,
they differ significantly from the standard (Peccei-Quinn, PQ)
invisible axion, whose mass range is expected in the milli-eV region.
This significant reduction of the axion mass is due to the suppression of the size of the instanton potential at the electroweak scale, rather than the hadronic one, which is responsible for inducing the coherent vacuum oscillations of the axion field. \\
According to this scenario, the field generated by the vacuum misalignment would be rolling down
very slowly towards the minimum of the non-perturbative instanton
potential emerging at the electroweak phase transition.\\
Various experimental studies have
significantly constrained the parameter space (axion mass and gauge
couplings) for the PQ axions
\cite{Arik:2008mq,Asztalos:2009yp,Duffy:2009ig}. The study of these
types of fields has also taken into account the possibility to evade
the experimental bounds \cite{Visinelli:2009zm}. These take the form of both an upper and a lower bound on the size of $f_a$,
the axion decay constant, which sets the scale of the misalignment
angle $\theta$, defined as the ratio of the axion field ($a$) over the
PQ scale $v_{PQ}$ ($v_{PQ}\sim f_a$).

While the agreement between $\Lambda$CDM and the observations is significant at most scales, at a small sub-galactic scale, 
corresponding to astrophysical distances relevant for the description of the stellar distributions ($\sim 10$ kpc), cold dark matter models predict an abundance of low-mass halos in excess of observations \cite{Hu:2000ke}. Difficulties in the description of this sub-galactic region have usually been attributed to an inaccurate modelling of its baryonic content, related to star formation mechanisms, supernova explosions and black hole activity in these regions, causing a redistribution of matter. Various suggestions to solve this discrepancy have been put forward, for instance the presence of warm dark matter of (WDM), whose free streaming could erase halos and sub-halos of low mass. As observed in \cite{Hu:2000ke} and later in \cite{Hui:2016ltb}, the resolution of these issues may require a cold dark matter component which is ultralight, in the $10^{-20}-10^{-22}$ eV range. \\
In the context of string theory, where massless moduli in the form of scalar and pseudoscalar fields abound at low energy, it is possible to consider the possibility that such states may be generated at the Planck scale, with their flat potentials slightly lifted in order to give rise to ultralight particles. 

\section{Non-supersymmetric scenarios}
The theoretical prediction for the mass range in which to locate a PQ axion is currently below the eV region in all the accepted formulations. We recall that the axion solution of the strong CP problem has been formulated according 
to two main scenarios, the KSVZ axion (or hadronic axion) 
and the DFSZ \cite{Dine:1981rt, Zhitnitsky:1980tq} axion, the latter introduced in a model which requires, in addition, a scalar sector with two Higgs doublets $H_u$ and $H_d$, besides the PQ complex scalar $\Phi$.\\
As already mentioned, in both cases the small axion mass is attributed to a vacuum misalignment mechanism generated by the structure of the QCD vacuum at the QCD phase transition, which determines a tilt in the flat PQ potential. 
The latter undergoes a symmetry breaking at a scale $v_{PQ}$, above the scales of inflation $H_I$ and of reheating $(T_R)$, and hence quite remote from the electroweak/confinement scales, though other relative locations for such scales are also possible.\\
In both scenarios the original symmetry can be broken by gravitational effects, and one must guarantee that the physical Goldstone mode $a(x)$ sits on the flat vacuum valley from the 
large $v_{PQ}$ scale down to $\Lambda_{QCD}$, when axion oscillations start. 
In the DFSZ solution, the axion emerges as a linear combination of the phases of the CP-odd sector and of $\Phi$ which are orthogonal to the hypercharge $(Y)$ and are fixed by the normalization of the kinetic term of the axion field $a$. 
The solution to the strong CP problem is then achieved by rendering the parameter of the $\theta$-vacuum dynamical, with the angle $\theta$ replaced by the axion field ($\theta\to 
 a/f_a$), with $f_a$ being the axion decay constant. 
 
\subsection{St\"uckelberg models with two-Higgs doublets }
The motivations in favour of such models are manifold, essentially motivated by the fact that anomalous $U(1)'s$ are ubiquitous in string compactifications and that the mechanism of anomaly cancellation in string theory finds a simple realization at field theory level in the form of a local anomaly action with a Stuckelberg field. Details on the origin of such models and on the corresponding anomaly action have been discussed in the past in various contexts, starting from past proposals of scenarios characterised by a low scale for gravity in the presence of large extra dimensions and of matter configurations assigned to insersecting branes \cite{Coriano:2005js}. \\
As already mentioned, the type of models investigated in the past have been  formulated around the TeV scale and discussed in detail in their various sectors  
\cite{Coriano:2006xh,Coriano:2007fw,Coriano:2007xg,Armillis:2008vp,Coriano:2008pg,Coriano:2009zh,Armillis:2008bg} \cite{Frampton:1996cc}.
 Their effective actions are characterised by the inclusion of dimension-5 operators, in order to restore the gauge invariance of the Lagrangian, broken by the anomalous gauge symmetry. Therefore, they are quite different from ordinary anomaly-free versions of the same theories. They include, beside one extra anomalous $U(1)_B$ symmetry, associated with a gauge boson $B_\mu$, a St\"uckelberg field $(b(x))$ and a set of 
scalars with a sufficiently wide CP odd sector in order to induce a mixing potential between the scalar ($H_u$, $H_d$) fields and the St\"uckelberg. This is obtained, for instance by requiring 
that the scalars carry a different charge under $U(1)_B$ symmetry, which allows the presence of a phase in the scalar sector  
   which induces the mixing between $b(x)$ and the CP-odd phases $\textrm{Im} H_u$ and 
$\textrm{Im} H_d$ 
\begin{equation}
\label{period}
V_{\not{P}\not{Q}}\sim \lambda_0 H_u^{\dagger}H_d e^{-i g_B (q_u-q_d)\frac{b}{2 M_{St}}} +\ldots
\end{equation}
which we will define more completely below. It is this mixing, generated at the 
scale at which the gauge symmetry is broken, with the two scalars $H_i$ $(i=u,d)$ acquiring a vev, that the shift symmetry of the $b(x)$ gets broken. Then, a physical component of the 
same field is generated $(\chi)$, which is directly coupled to the anomalies of the model in the form
\beqa
b &=& O_{33}^{\chi} \chi + G's.      
\label{rot}
\eeqa
In \eqref{rot} the $G$'s denote the Goldstones of the CP-odd sector of the model. One recognizes in $\chi(x)$ a physical axion, which couples to the gauge fields as an ordinary axion   
\begin{equation} 
\frac{b(x)}{M_{St}} F\tilde{F}  \to \frac{\chi(x)}{M_{St}} F\tilde{F}.
\end{equation}
Notice that $\chi(x)$, differently from $b(x)$, is gauge invariant and will appear as the physical phase of the periodic potential \eqref{period}, as for an ordinary PQ axion. In this construction, 1) the coupling of the physical axion to the anomaly and 2) its mass, are not controlled by a single scale $(f_a)$ as in PQ, but by two unrelated scales. Therefore, this scenario allows a far wider parameter space where one could look for such particles \cite{Coriano:2006xh,Coriano:2007xg,Coriano:2009zh}.\\
 So far, only two complete models have been put forward for a consistent analysis of these types of particles, the first one non-supersymmetric  \cite{Coriano:2005js} and a second one supersymmetric \cite{Coriano:2008xa}. In the supersymmetric case, the St\"uckelberg turns into a supermultiplet, with an axion, a saxion and an axino. $\chi$ defines a typical axion-like particle, introduced on the basis of a well-defined gauge structure rather than on purely phenomenological grounds. \\ 
\subsection{Models with $M_S$ in the  TeV/multi TeV range} 
\begin{figure}[t]
\begin{align}
S_{eff}=&S_0+
\begin{minipage}[c]{70pt}
\includegraphics[scale=.5]{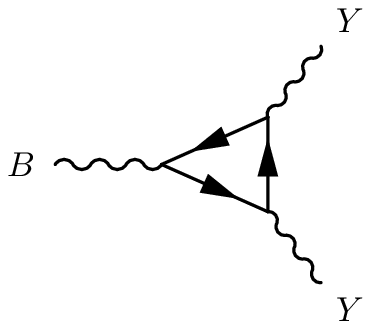}
\end{minipage}+
\begin{minipage}[c]{70pt}
\includegraphics[scale=.5]{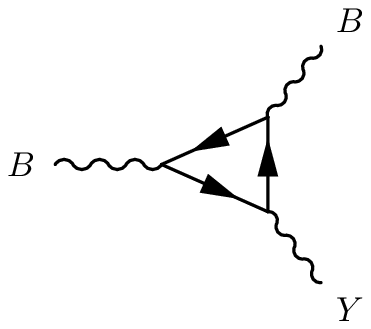}
\end{minipage}+
\begin{minipage}[c]{70pt}
\includegraphics[scale=.5]{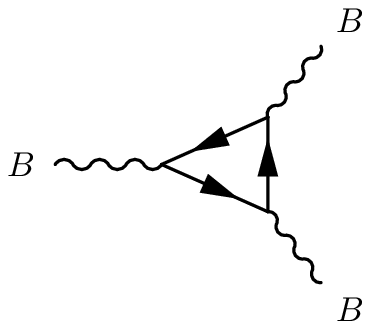}
\end{minipage}+
\begin{minipage}[c]{70pt}
\includegraphics[scale=.5]{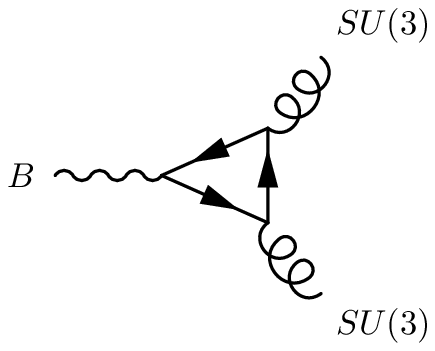}
\end{minipage}+
\begin{minipage}[c]{70pt}
\includegraphics[scale=.5]{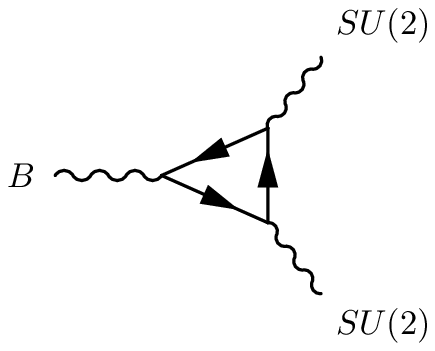}
\end{minipage}+\nonumber\\
&\nonumber\\
&
\begin{minipage}[c]{70pt}
\includegraphics[scale=.5]{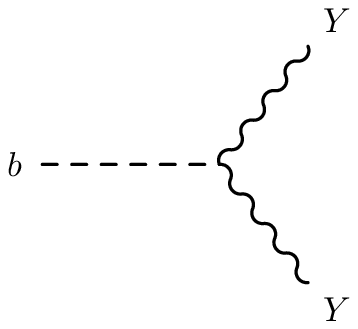}
\end{minipage}+
\begin{minipage}[c]{70pt}
\includegraphics[scale=.5]{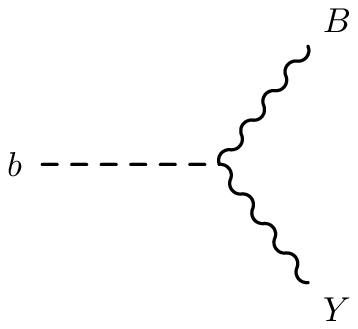}
\end{minipage}+
\begin{minipage}[c]{70pt}
\includegraphics[scale=.5]{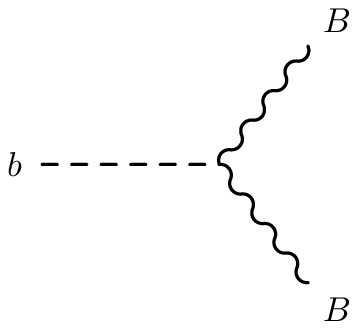}
\end{minipage}+
\begin{minipage}[c]{70pt}
\includegraphics[scale=.5]{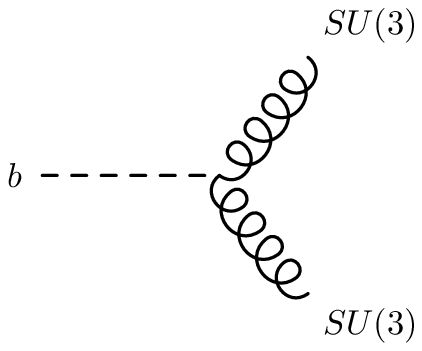}
\end{minipage}+
\begin{minipage}[c]{70pt}
\includegraphics[scale=.5]{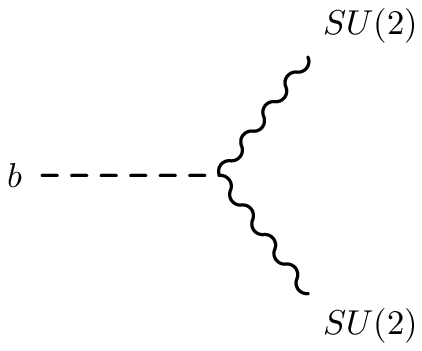}
\end{minipage}\nonumber
\end{align}
\caption{The 1PI effective action for a typical low scale model with the addition of one extra anomalous $U(1)_B$ to the Standard Model. Shown are the one-loop trilinear anomalous interactions and the corresponding counterterms, involving the St\"uckelberg field $b(x)$.} 
\label{fig:lagrangian}
\label{eaction}
\end{figure}

Originally  such models have been formulated in order to define a consistent framework for the search of anomalous gauge interactions and of extra neutral currents (extra Z prime's) \cite{Armillis:2008vp,Armillis:2007tb} at the LHC, with a St\"uckelberg scale $M_{St}$ assumed around the TeV/multi TeV range. Notice that $M_{ST}$ is of the order of the mass of the anomalous gauge boson $B_\mu$, which appears at tree level. In fact the St\"uckelberg field $b(x)$ acquires a gauge invariant kinetic term by a direct mixing with the $B_\mu$ in the form

\begin{equation}
\label{deff}
\mathcal{L}_{St}=\frac{1}{2}(\partial_\mu b - M_{St} B_\mu)^2
\end{equation}
with the $b(x)$ charged only under the $U(1)_B$ anomalous symmetry 
\begin{equation} 
b(x)\to b(x) + M_{St} \theta_B(x) \qquad B_\mu(x) \to B_\mu(x) + \partial_\mu \theta_B(x).
\end{equation} 
In all these models the effective action has the structure given by
\beqa
{\mathcal S} &=&   {\mathcal S}_0 + {\mathcal S}_{Yuk} +{\mathcal S}_{an} + {\mathcal S}_{WZ} 
\label{defining}
\eeqa
where ${\mathcal S}_0$ is the classical SM action, with the inclusion of one extra abelian symmetry, $U(1)_B$. The same structure will also characterize other, more complex, realizations, and the extra symmetry can be attached at any scale, in principle.\\
In \eqref{defining} we have denoted with ${\mathcal S}_{Yuk}$ the Yukawa interactions, 
which will generate a coupling of the gauge invariant field $\chi$ to the fermion sector. It contains the usual gauge degrees of freedom of the Standard Model plus the extra anomalous gauge boson $B$ which is already massive before electroweak symmetry breaking, as clear from \eqref{deff}. The coupling of $b(x)$ to the anomaly is contained in $\mathcal{S}_{WZ}$.\\
We show the structure of the 1-particle irreducible effective action in Fig. \ref{eaction}. In this case, consider a 2-Higgs doublet model for definiteness, which will set the ground for more complex extensions that we will address in a final section. We consider 
a $SU(3)_c\times SU(2)_w\times U(1)_Y\times U(1)_B$ gauge symmetry model, characterized by an action $\mathcal{S}_0$, corresponding to the first contribution shown in Fig.~\ref{eaction}, 
plus one loop corrections which are anomalous and break gauge invariance. Any insertion of a $U(1)_B$ gauge current in the trilinear fermion vertices will generate anomalies which are cancelled by the operators introduced in $\mathcal{S}_{Yuk}$. \\
\begin{table}[t]
\begin{center}
\begin{tabular}{|c|c|c|c|c|c|c|}
\hline
$ f $ & $Q$ &  $ u_R $ &  $ d_R $ & $ L $ & $e_R$ \\
\hline \hline
$q^B$  &  $q^B_Q$  & $q^B_{u_R}$  &  $q^B_{d_R}$ & $q^B_L$ & $q^B_{e_R}$ \\ \hline
\end{tabular}
\end{center}
\centering
\renewcommand{\arraystretch}{1.2}
\begin{tabular}{|c|c|c|c|c|}\hline
$f$ & $SU(3)^{}_C$ & $SU(2)^{}_L$ & $U(1)^{}_Y$ & $U(1)^{}_B$ \\ \hline \hline
$Q$ &  3 & 2 & $ 1/6$ & $q_Q^B$\\
$u_R$ &  3 & 1 & $ 2/3$ & $q_Q^B+q^B_{u}$\\
$d_R$ &  3 & 1 & $ -1/3$ & $q_Q^B-q^B_{d}$\\
$L$ &  1 & 2 & $ -1/2$ & $q^B_L $\\ 
$e_R$ &  1 & 1 & $-1$ & $ q^B_L - q^B_{d}$\\ 
\hline
$H^{}_u$ &  1 & 2 & $1/2$ & $ q^B_u $\\ 
$H^{}_d$ &  1 & 2 & $1/2$ & $ q^B_d  $\\ 
\hline
\end{tabular}
\caption{\small Charges of the fermion and of the scalar fields \label{solve_q}}
\label{Table}
\end{table}
In the last line of the same figure are shown the $(b/M_{St}) F\tilde{F}$ Wess-Zumino counterterms needed for restoring gauge invariance. Table \ref{Table} shows the charge assignments of the fermion spectrum of the model, where we have indicated by $q$ the charges for a single generation. The two Higgs fields carry different charges under $U(1)_B$, which allows to extend the ordinary scalar potential of the 
 two-Higgs doublet model by an extra contribution with the inclusion of the real St\"uckelberg field $b(x)$. This will be periodic in the axi-Higgs $\chi$, after the two Higgses, here denoted as $H_u$ and $H_d$, acquire a vev. 
Specifically, $ q^B_L, q^B_Q$ denote the charges of the left-handed lepton doublet $(L)$ and of the quark doublet $(Q)$ respectively, while $q^B_{u_r},q^B_{d_r}, q^B_{e_R}$ are the charges of the 
right-handed $SU(2)$ singlets (quarks and leptons). We denote by $\Delta q^B=q^B_u - q^B_d$ the difference between the two charges of the up and down Higgses $(q^B_u, q^B_d)$ respectively, chosen to be non-zero.
The trilinear anomalous gauge interactions induced by the anomalous $U(1)$ and the relative counterterms, which are all parts of the 1-loop effective action, are illustrated in Fig. \ref{fig:lagrangian}. 

\subsection{The scalar sector}
The simplest realization of such a model is generated by including in the scalar sector two Higgs doublets, in a potential $V_{P Q}(H_u, H_d)$. The gauge symmetry is that of the SM times an extra abelian group, $U(1)_B$. The St\"uckelberg field appears in a second term $V_{\slashed{P}\slashed{Q}}(H_u,H_d,b)$ - or $V^\prime$ (PQ breaking)  
\cite{Coriano:2005js} - as a phase, which is allowed by the symmetry and mixes the Higgs sector with the St\"uckelberg axion $b$ \cite{Coriano:2005js}, 
\begin{equation}
\label{pot}
V=V_{PQ}(H_u,H_d) + V_{\slashed{P}\slashed{Q}}(H_u,H_d,b).
\end{equation}

The appearance of the physical axion in the spectrum of the model takes place after spontaneous symmetry breaking of the two Higgs fields, and   
the phase-dependent terms - here assumed to be of non-perturbative origin and generated at a phase transition - induce a tiny curvature on the scalar potential at the bottom of the vacuum valley. The mixing induced in the CP-odd sector generates a mass for
of a linear combination of the St\"uckelberg field $b$ and the CP-odd Goldstones, indicated as $\chi$, the physical axion. From (\ref{pot}) we have as first term
\beqa
V_{PQ}&=&\mu_u^2 H_u^{\dagger}H_u+\mu_d^2 H_d^{\dagger}H_d+\lambda_{uu}(H_u^{\dagger}H_u)^2
+\lambda_{dd}(H_d^{\dagger}H_d)^2-2 \lambda_{ud}(H_u^{\dagger}H_u)(H_d^{\dagger}H_d)
+2\lambda^{\prime}_{ud}\vert H_u^T \tau_2 H_d\vert^2 \nonumber \\
\eeqa
typical of a two-Higgs doublet model, to which we add a second PQ breaking term
\beqa
V_{\slashed{P}\slashed{Q}}&=&\lambda_0(H_u^{\dagger}H_d e^{-i g_B (q_u-q_d)\frac{b}{2 M}})+
\lambda_1(H_u^{\dagger}H_d e^{-i g_B (q_u-q_d)\frac{b}{2 M}})^2+\lambda_2(H_u^{\dagger}H_u)(H_u^{\dagger}H_d e^{-i g_B (q_u-q_d)\frac{b}{2M}})+\nonumber\\
&&\lambda_3(H_d^{\dagger}H_d)(H_u^{\dagger}H_d e^{-i g_B (q_u-q_d)\frac{b}{2 M}})+\textrm{h.c.}
\eeqa
These terms are allowed by the symmetry of the 
model and are parameterized by one dimensionful  ($\lambda_0$) and three dimensionless couplings ($\lambda_1,\lambda_2,\lambda_3$).  
As already mentioned, their values are weighted by an exponential factor containing as a suppression proportional to the instanton action \cite{Coriano:2010py}. In the equations below we will rescale 
$\lambda_0$ by the electroweak scale $v=\sqrt{v_u^2 + v_d^2}$  ($\lambda_0 \equiv \bar{\lambda}_0 v$), to obtain a homogeneous expression for the mass of $\chi$ 
as a function of the relevant scales of the model. These are, besides the electroweak vev $v$, the St\"uckelberg mass $M_{St}$ and the anomalous gauge coupling of the $U(1)_B$, $g_B$. Such gauge coupling is relevant in the analysis of possible extra anomalous neutral currents in collider searches, but does not play any significant role otherwise, especially if $M_{St}$ is assumed to be very large. This is the most interesting case if we consider scenarios with an ultralight axion.  

\subsection{The potential for a generic St\"uckelberg mass} 
 
The physical axion $\chi$ emerges as a linear combination of the phases of the various complex scalars with the $b$ field. 
To illustrate the appearance of a physical direction in the phase of the extra potential, we focus our attention just on its CP-odd sector, 
which is the only one that is relevant for our discussion.  The expansion of this potential around the electroweak vacuum is given by the parameterization 
\beqa
H_u=\left(
\begin{tabular}{c}
$H_u^+$\\
$v_u+H_u^0$
\end{tabular}
\right)
\hspace{1cm}
H_d=\left(
\begin{tabular}{c}
$H_d^+$\\
$v_d+H_d^0$
\end{tabular}
\right),
\eeqa
where $v_u$ and $v_d$ are the two vevs of the Higgs fields.
This potential is characterized by two null eigenvalues corresponding to two neutral Nambu-Goldstone modes $(G_0^1,G_0^2)$,
and an eigenvalue corresponding to a massive state with an axion component ($\chi$). In the 
$(\textrm{Im}H_d^0,\textrm{Im} H_u^0, b)$ CP-odd basis this is given by

\begin{align}
\chi&=\frac{1}{\sqrt{g_B^2 (q_d-q_u)^2 v_u^2 v_d^2+2M^2(v_d^2 + v_u^2)}}
\left(\sqrt{2} M v_u,-\sqrt{2} M v_d, g_B (q_d-q_u) v_d v_u\right)
\end{align}
and we indicate with $O^{\chi}$ the orthogonal matrix which allows to rotate them to the physical basis
\beqa
\begin{pmatrix}
\label{ochi}
G_0^1 \cr
G_0^2 \cr
\chi
\end{pmatrix}
= O^\chi
\begin{pmatrix}
\textrm{Im}H^0_d \cr
\textrm{Im}H^0_u \cr
b
\end{pmatrix},
\eeqa

$\chi$ inherits WZ interaction since $b$ can be related to the physical axion $\chi$ and to the Nambu-Goldstone modes via this matrix as
\beqa
b &=&  O_{13}^{\chi} G_0^1 + O_{23}^{\chi} G_0^2 + O_{33}^{\chi} \chi ,       
\label{rot12}
\eeqa
or, conversely,
\beqa
\chi &=& O_{31}^{\chi} \textrm{Im}H_d + O_{32}^{\chi} \textrm{Im}H_u + O_{33}^{\chi} b.        
\eeqa
One can show that the rotation of $b$ into the physical axion $\chi$ involves a factor $O_{33}^{\chi}$ which is of order $v/M_{St}$.
This implies that $\chi$ inherits from $b$ an interaction with the gauge fields which is suppressed by a scale $M_{St}^2/v$. This scale is the product of two contributions: a $1/M_{St}$ suppression coming from the original Wess-Zumino counterterm of the Lagrangian ($b/M_{St} F\tilde{F}$), and of a factor $v/M_{St}$, obtained by the projection of $b$ into $\chi$ due to $O_\chi$.\\
The final expression of the coupling of the axi-Higgs to the photon $g_{\chi\gamma\gamma}\chi F_\gamma\tilde{F_\gamma}$, 
is defined by a combination of matrix elements of the rotation matrices $O^A$, which defines the gauge eigenstates, and of $O^\chi$. Defining $g^2=g_2^2 + g_Y^2$, in terms of the $SU(2)$ and $U(1)_Y$ couplings, one obtains for such a coupling an expression in terms of the left and right $(L,R)$ gauge charges of the fermions $(f)$, because of the anomaly
\beq
g^{\chi}_{\g \g}\,= \frac{g_B g_Y^2 g_2^2}{32 \pi^2 M_{St} g^2} O^{\chi}_{3\,3}\sum_f\left(- q^B_{f\,L} +q^B_{f \,R} \left(q^Y_{f \,R}\right)^2 - q^B_{f \,L} \left(q^Y_{f \,L}\right)^2\right).
 \label{gchi}
 \eeq 

Notice that this expression is cubic in the gauge coupling constants, since factors such as $g_2/g$ and $g_Y/g$ are mixing angles, while the factor $1/\pi^2$ originates from the one-loop suppression in the anomaly diagram. Therefore  one obtains a general behaviour for $g^{\chi}_{\g \g}$ which is of order $O(g^3 v/M_S^2)$, while the charges are, in general, of order unity. 
\subsection{The periodic potential as a function of the physical axion  }
The results above can be reobtained by looking at the phases of the extra potential, proving the periodicity of such potential in the field variable $\chi(x)$, which is the physical axion. This second approach shows quite directly that$\chi(x)$, extracted from $b(x)$, is gauge invariant. In fact, if we choose a polar parametrization of the neutral components in the broken phase 
\beqa
H_u^0=\frac{1}{\sqrt{2}}\left(\sqrt{2}v_u + \rho_u^0(x) \right) e^{i\frac{F_u^0(x)}{\sqrt{2}v_u}}
\hspace{1cm}
H_d^0=\frac{1}{\sqrt{2}}\left(\sqrt{2}v_d + \rho_d^0(x) \right) e^{i\frac{F_d^0(x)}{\sqrt{2}v_d}},
\eeqa
where we have introduced the two phases $F_u$ and $F_d$ of the two neutral Higgs fields, the information on the periodicity is  obtained by linearly combining all the phases of $V'$ as 
\beqa
\theta(x)\equiv\frac{g_B (q_d-q_u)}{2 M_{St}}b(x)-\frac{1}{\sqrt{2}v_u} F_u^0(x) +\frac{1}{\sqrt{2}v_d} F_d^0(x).
\eeqa
Using the matrix $O^{\chi}$ to rotate on the physical basis of the CP-odd scalar sector, the phase describing the periodicity of the potential turns out to be proportional to the physical axion $\chi$, modulo a dimensionful constant ($\sigma_\chi$)
\beqa
\theta(x)\equiv \frac{\chi(x)}{\sigma_\chi},
\label{theta}
\eeqa
where we have defined
\beq
\sigma_\chi\equiv\frac{2  v_u v_d M_{St}}{\sqrt{g_B^2 (q_d-q_u)^2 v_d^2 v_u^2 +2 M^2 (v_d^2+v_u^2)}}.
\eeq
The constant $\sigma_\chi$, in our case, takes the same role of $f_a$ in the PQ case, where the angle of misalignment is identified by the ratio $a(x)/f_a$, with $a(x)$ being the PQ axion. \\
As already mentioned, the re-analysis of the $V'$ potential is particularly useful for proving the gauge invariance of $\chi$ under a $U(1)_B$ infinitesimal gauge transformation with gauge parameter $\alpha_B(x)$.\\
We refer to \cite{Coriano:2010py} for more details on this point. After spontaneous symmetry breaking induced by the two doublets, the periodic potential takes the form
\begin{align}
V^\prime=& 4 v_u v_d
\left(\lambda_2 v_d^2+\lambda_3 v_u^2+\lambda_0\right) \cos\left(\frac{\chi}{\sigma_\chi}\right) + 2 \lambda_1 v_u^2 v_d^2 \cos\left(2\frac{\chi}{\sigma_\chi}\right),
\label{extrap}
\end{align}
with a mass for the physical axion $\chi$  given by
\beqa
m_{\chi}^2=\frac{2 v_u v_d}{\sigma^2_\chi}\left(\bar{\lambda}_0 v^2 +\lambda_2 v_d^2 +\lambda_3 v_u^2+4 \lambda_1 v_u v_d\right) 
\approx \lambda v^2.
\label{axionmass}
\eeqa
Eq. \eqref{axionmass} shows that the size of the potential is driven by the combination of non-perturbative effects, parameterized by the exponentially small constants 
$(\bar{\lambda_0}, \lambda_1,\lambda_2,\lambda_3)$, with the electroweak vevs of the two Higgses. 
One point that needs to be stressed is that at the electroweak epoch the angle of misalignment generated by the extra potential is parameterized by $\chi/\sigma_\chi$, while the interaction of the physical axion with the gauge fields is suppressed by $M_{St}^2/v$. This feature sets a difference compared to the PQ case, where both scales are identified, and given by the axion decay constant $f_a$. \\
Notice that if we assume that $M_{St}$ is identified with the breaking scale $v$, then the suppression of the coupling of 
$\chi$ to the divergence $F\tilde{F}$ is simply controlled by the same scale $M_{St}$. In a final section we are indeed going to do so, and choose $M_{St}$ of the order of the GUT scale. This is sufficient for predicting an ultralight axion-like particle. Obviously, the scalar sector, in this case, is far more involved, due to the large gauge structure and to the proliferation of scalars. However, the results obtained in the simplest 2-Higgs doublet model can be extended to such general models. 

\section{Supersymmetric extensions}
The study of models with an anomalous $U(1)$ interaction in the spectrum has proceeded in a rather direct way \cite{Fucito:2008ai,Anastasopoulos:2008jt}, with the goal of formulating workable models which could be tested at ordinary colliders, both with and without supersymmetry. \\ 
However, anomalous abelian extensions of the MSSM do not allow a physical axion-like particle for the absence of Higgs-axion mixing.
The NMSSM (next-to-minimal supersymmetric standard model), enlarged by an extra anomalous $U(1)$, is the first model in which the construction reviewed above can be consistently formulated. In the literature on supersymmetry phenomenology, anomaly-free models of this type are referred to as USSM-like, i.e. supersymmetric Standard Models (SSM) with one extra $U(1)$ \cite{Cvetic:1997ky}). In \cite{Coriano:2010ws,Coriano:2008xa} the inclusion of an extra (anomalous) $U(1)$ led to the formulation of a model, called "the USSM-A", in order to emphasize its anomalous content (-A). 
The gauge symmetry of such models is the same of the Standard Model, with the inclusion of the extra $U(1)$ and of two Higgs doublets superfields.
The superpotential is the same of the NMSSM, with the inclusion of one extra singlet scalar superfield $\hat{S}$. Also in this case, as in the non-supersymmetric versions, the operatorial content of the defining Lagrangian that we have described above is not modified, but now the St\"uckelberg field is promoted to a supermultiplet, with several components.\\
We will focus our discussion on the axion/saxion Lagrangian, which illustrates the new features of this supermultiplet and it is defined as
\begin{align} {\cal L}_{axion/saxion}={\cal L}_{St} +{\cal L}_{WZ},
\end{align} where ${\cal L}_{St}$ is the supersymmetric version of the
St\"uckelberg mass term \cite{Kors:2004ri}, while ${\cal L}_{WZ}$
denotes the WZ counterterms responsible for the axion-like nature of
the pseudoscalar $b$. Specifically,
\begin{align} 
{\cal L}_{St}&=\frac{1}{2}\int d^{4}\theta (\hat{{\bf
b}} +\hat{{\bf b}}^{\dagger}+\sqrt{2} M_{St} \hat{B})^{2}\nonumber\\ {\cal
L}_{WZ} &= -\frac{1}{2}\int d^{4}\theta\left\lbrace
\left[ \frac{c_{G}}{M_{St}} \,\textrm{Tr}({\cal G} {\cal G})\hat{{\bf b}}
+ \frac{c_{W}}{M_{St}} \,\textrm{Tr}(W W)\hat{{\bf b}}
\right.\right.  \nonumber\\
&\left.\left.+\frac{c_{Y}}{M_{St}}\hat{{\bf b}}W^{Y}_{\alpha}
W^{Y,\alpha} +\frac{c_{B}}{M_{St}}\hat{{\bf
b}}W^{B}_{\alpha}W^{B,\alpha}+\frac{c_{YB}}{M_{St}}\hat{{\bf
b}}W^{Y}_{\alpha}W^{B,\alpha}\right] \delta(\bar{\theta}^{2})
+h.c.\right\rbrace,
\end{align} 
where we have denoted with $\cal{G}$ the supersymmetric
field-strength of $SU(3)_c$, with $W$ the supersymmetric
field-strength of $SU(2)$, with $W^{Y}$ and with $W^{B}$ the
supersymmetric field-strength of $U(1)_{Y}$ and of the anomalous abelian symmetry $U(1)_{B}$
respectively. 
The St\"uckelberg supermultiplet $\hat{\bf b}$ takes the form
\begin{align} 
\hat{{\bf b}}= b + \sqrt{2}\theta\psi_{{\bf b}}
- i\theta\sigma^{\mu}\bar{\theta}\partial_{\mu} b
+ \frac{i}{\sqrt{2}}\theta\theta\bar{\theta}\bar{\sigma}^{\mu}\partial_{\mu}\psi_{{\bf
b}} -\frac{1}{4}\theta\theta\bar{\theta}\bar{\theta}\Box b 
- \theta\theta F_{{\bf b}},
\end{align} 
and contains the St\"uckelberg axion (now a complex $b$ field)
and its supersymmetric partner, referred to as the axino ($\psi_{\bf
b}$), which combines with the neutral gauginos and higgsinos to
generate the neutralinos of the model. Details on the notation for the
superfields components can be found in Tab.~\ref{superfieldcomp}.  We have denoted with 
$\lambda_B$ and $\lambda_Y$ the two gauginos of the two vector superfields $(\hat{B}, \hat{Y})$ corresponding to the anomalous $U(1)_B$ and the hypercharge vector multiplets.  
The superfield $\hat{S}$ has as components the scalar $S$, simply called "the singlet"  and its supersymmetric partner, the singlino, 
denoted as $\tilde{S}$. 
\begin{table}
\begin{center}
\begin{tabular}{|l||c|c|c|} \hline Superfield & Bosonic & Fermionic &
Auxiliary \\ \hline $\hat{\bf b}(x,\theta,\bar{\theta})$& $b(x)$ &
$\psi_{\bf b}(x)$ & $F_{\bf b}(x)$ \\ $\hat{S}(x,\theta,\bar{\theta})$
& $S(x)$ & $\tilde{S}(x)$ & $F_{S}(x)$ \\
$\hat{L}(x,\theta,\bar{\theta})$ & $\tilde{L}(x)$ & $L(x)$ &
$F_{L}(x)$\\ $\hat{R}(x,\theta,\bar{\theta})$ & $\tilde{R}(x)$ &
$\bar{R}(x)$ & $F_{R}(x)$ \\ $\hat{Q}(x,\theta,\bar{\theta})$ &
$\tilde{Q}(x)$ & $Q(x)$ & $F_{Q}(x)$ \\
$\hat{U}_{R}(x,\theta,\bar{\theta})$& $\tilde{U}_R(x)$ &
$\bar{U}_R(x)$ & $F_{U_R}(x)$ \\ $\hat{D}_{R}(x,\theta,\bar{\theta})$&
$\tilde{D}_R(x)$ & $\bar{D}_R(x)$ & $F_{D_R}(x)$ \\
$\hat{H}_{1}(x,\theta,\bar{\theta})$& $H_1(x)$ & $\tilde{H_1}(x)$ &
$F_{H_1}(x)$ \\ $\hat{H}_{2}(x,\theta,\bar{\theta})$& $H_2(x)$ &
$\tilde{H_2}(x)$ & $F_{H_2}(x)$ \\ $\hat{B}(x,\theta,\bar{\theta})$ &
$B_{\mu}(x)$ & $\lambda_{B}(x)$ & $D_B(x)$ \\
$\hat{Y}(x,\theta,\bar{\theta})$ & $A^{Y}_{\mu}(x)$ & $\lambda_{Y}(x)$
& $D_Y(x)$ \\ $\hat{W}^{i}(x,\theta,\bar{\theta})$& $W^{i}_{\mu}(x)$ &
$\lambda_{W^{i}}(x)$ & $D_{W^{i}}(x)$ \\
$\hat{G}^{a}(x,\theta,\bar{\theta})$& $G^{a}_{\mu}(x)$ &
$\lambda_{g^a}(x),\bar{\lambda}_{g^a}(x)$ & $D_{G^{a}}(x)$ \\ \hline
\end{tabular}
\end{center}
\caption{Superfields and their components.}
\label{superfieldcomp}
\end{table}

The Lagrangian ${\mathcal L}_{St}$ is invariant under
the $U(1)_B$ gauge transformations
\begin{align} \delta_B
\hat{B}=&\hat{\Lambda}+\hat{\Lambda}^\dagger\nonumber\\ \delta_B
\hat{\bf{b}}=& - 2 M_{St} \hat{\Lambda}
\end{align} where $\hat{\Lambda}$ is an arbitrary chiral
superfield. So the scalar component of $\hat{\bf{b}}$, that consists
of the saxion and the axion field, shifts under a $U(1)_B$ gauge
transformation.\\ The coefficients $c_I\equiv(c_G,c_W, c_Y, c_B,
c_{YB})$ are constants, fixed by the conditions of gauge
invariance. They are functions of the free charges  of the model and are related to the cancellation of the anomalies $\left\{U(1d )_{B}^3\right\}$,
$\left\{U(1)_{B},U(1)_{Y}^2\right\}$,
$\left\{U(1)_{B}^2,U(1)_{Y}\right\}$,
$\left\{U(1)_{B},SU(2)^2\right\}$,
$\left\{U(1)_{B},SU(3)^2\right\}$. \\
The superpotential is chosen of the form
\begin{align} {\cal W} = \lambda
\hat{S}\hat{H}_{1}\cdot\hat{H}_{2}+y_{e}\hat{H}_{1}
\cdot\hat{L}\hat{R}+y_{d}\hat{H}_{1}\cdot\hat{Q}\hat{D}_{R}
+y_{u}\hat{H}_{2}\cdot\hat{Q}\hat{U}_{R}.
\label{supi}
\end{align} 
This superpotential, as shown in
\cite{Coriano:2008xa,Coriano:2008aw}, allows a physical axion in the
spectrum. The gauge sector and the the soft breaking terms, in the form
of scalar mass terms (SMT), are identical to those of the USSM \cite{Cvetic:1997ky}
\begin{align} 
&{\cal L}_{gauge}=\frac{1}{4}\int{d^{4}\theta\left[{\cal
G}^{\alpha}{\cal
G}_{\alpha}+W^{\alpha}W_{\alpha}+W^{Y\alpha}W^{Y}_{\alpha}
+W^{B\alpha}W^{B}_{\alpha}\right]\delta^{2}(\bar{\theta})+h.c.}
\nonumber\\ 
&{\cal L}_{SMT}=-\int
d^{4}\theta\,\delta^{4}(\theta,\bar{\theta})\,[M^{2}_{L}\hat{L}^{\dagger}\hat{L}
+m^{2}_{R}\hat{R}^{\dagger}\hat{R}+M^{2}_{Q}\hat{Q}^{\dagger}\hat{Q}
+m^{2}_{U}\hat{U}_{R}^{\dagger}\hat{U}_{R}+m^{2}_{D}\hat{D}_{R}^{\dagger}\hat{D}_{R}
\nonumber\\ &\hspace{2.5cm}+m_{1}^{2}\hat{H}_{1}^{\dagger}\hat{H}_{1}
+m_{2}^{2}\hat{H}_{2}^{\dagger}\hat{H}_{2}+m_{S}^{2}\hat{S}^{\dagger}\hat{S}
+(a_{\lambda}\hat{S}\hat{H}_{1}\cdot\hat{H}_{2}+h.c.)+(a_{e}\hat{H}_{1}\cdot\hat{L}\hat{R}+h.c.)
\nonumber\\
&\hspace{2.5cm}+(a_{d}\hat{H}_{1}\cdot\hat{Q}\hat{D}_{R}+h.c.)
+(a_{u}\hat{H}_{2}\cdot\hat{Q}\hat{U}_{R}+h.c.)].
\end{align} 
As usual, $M_{L},M_{Q},m_R,m_{U_R},m_{D_R},m_1,m_2,m_S$
are the mass parameters of the explicit supersymmetry breaking, while
$a_e,a_{\lambda},a_u,a_d$ are dimensionful coefficients.  The soft
breaking due to gaugino mass terms (GMT) now include a mixing mass
parameter $M_{YB}$
\begin{align} 
&{\cal L}_{GMT}=\int d^{4} \theta \left[ \frac{1}{2}
\left(M_{G}{\cal G}^{\alpha}{\cal G}_{\alpha} + M_{w}W^{\alpha
}W_{\alpha} + M_YW^{Y\alpha} W^{Y}_{\alpha} + M_B W^{B\alpha}
W^{B}_{\alpha} \right. \right. \nonumber \\ &\qquad\qquad\qquad\qquad
\left. \left. + M_{Y B} W^{Y\alpha} W^{B}_{\alpha} \right)
+h.c.\right] \delta^{4}(\theta,\bar{\theta}).
\end{align} 

\begin{table}
\begin{center}
\begin{tabular}{|l||c|c|c|c|} \hline Superfields &SU(3)& SU(2)&
$U(1)_{Y}$ & $U(1)_{B}$\\ \hline $\hat{\bf b}(x,\theta,\bar{\theta})$&
{\bf 1} & {\bf 1} & 0 & $ s $\\ $\hat{S}(x,\theta,\bar{\theta})$& {\bf
1} & {\bf 1} & 0 & $B_{S}$\\ $\hat{L}(x,\theta,\bar{\theta})$& {\bf 1}
& {\bf 2} & -1/2 & $B_{L}$\\ $\hat{R}(x,\theta,\bar{\theta})$& {\bf 1}
& {\bf 1} & 1 & $B_{R}$\\ $\hat{Q}(x,\theta,\bar{\theta})$& {\bf 3} &
{\bf 2} & 1/6 & $B_{Q}$\\ $\hat{U}_{R}(x,\theta,\bar{\theta})$&
$\bar{{\bf 3}}$ & {\bf 1} & -2/3 & $B_{U_R}$\\
$\hat{D}_{R}(x,\theta,\bar{\theta})$& $\bar{{\bf 3}}$ & {\bf 1} & +1/3
& $B_{D_R}$\\ $\hat{H}_{1}(x,\theta,\bar{\theta})$& {\bf 1} & {\bf 2}
& -1/2 & $B_{H_{1}}$\\ $\hat{H}_{2}(x,\theta,\bar{\theta})$& {\bf 1} &
{\bf 2} & 1/2 & $B_{H_{2}}$\\ \hline
\end{tabular}
\label{cariche}
\end{center}
\caption{Charge assignment of the USSM-A model}
\label{fieldcontent}
\end{table} 
We will focus our attention on the main features of the St\"uckelberg sector, which is the most relevant for our purposes. 
\subsection{Axions and saxions}
The contributions of the axion and saxion to the total Lagrangian
are derived from the combination of the St\"uckelberg and Wess-Zumino
terms. The corresponding Lagrangian expressed in terms of
component fields is quite lengthy, but it exhibits a structure which is quite close to the one we have underlined in the non-supersymmetric case
\begin{align} 
{\cal L}_{axion/saxion}\equiv{\cal
L}_{St} + {\cal L}_{WZ} 
\end{align}
and contains a mixing among the $D$-terms
which is rather peculiar. The off-shell Lagrangian is given by
\begin{align} 
&{\cal L}_{axion/saxion}=\frac{1}{2}\left( \partial_{\mu}\textrm{Im}\,b +
M_{st}B_{\mu}\right)^{2}+\frac{1}{2}\partial_\mu\textrm{Re}b\,\partial^\mu\textrm{Re}b
+\frac{i}{2}\psi_{\bf b}\sigma^{\mu}\partial_{\mu}\bar{\psi_{\bf b}}
+\frac{i}{2}\bar{\psi_{\bf
b}}\bar{\sigma}^{\mu}\partial_{\mu}\psi_{\bf b} +F_{\bf b}^{\dagger}F_{\bf
b}+ {\cal L}_{axion, i}
\end{align}
where the expression of ${\cal L}_{axion, i}$ is quite lengthy and can be found in the original analysis \cite{Coriano:2010ws}. From the terms explictly reported above, one recognizes immediately that $\textrm{Re} b$ is the kinetic term of an ordinary scalar, while $\textrm{Im} b$ is the candidate to take the role of an ordinary axion. $F_{\bf b}$ is an auxiliary field which can be expressed in terms of the gaugino components 
\begin{align} 
F_{\bf b}=
-\frac{1}{16}\frac{c_{G}}{M_{St}}\bar{\lambda}^a_{G}\bar{\lambda}^a_{G}
-\frac{1}{16}\frac{c_{W}}{M_{St}}\bar{\lambda}^i_{W}\bar{\lambda}^i_{W}
-\frac{1}{2}\frac{c_{Y}}{M_{St}}\bar{\lambda}_{Y}\bar{\lambda}_{Y}
-\frac{1}{2}\frac{c_{B}}{M_{St}}\bar{\lambda}_{B}\bar{\lambda}_{B}
+\frac{1}{2}\frac{c_{YB}}{M_{St}}\bar{\lambda}_{Y}\bar{\lambda}_{B}.
\end{align}

There are some peculiar features of this model. While $\textrm{Im}\,
b$ may appear in the CP-odd part of the scalar sector and undergoes
mixing with the Higgs sector, its real part, $\textrm{Re}\, b$, the
saxion (or scalar axion) before the breaking of the gauge symmetry, has a mass exactly 
equal to the St\"uckelberg mass, as expected from supersymmetry. 
Indeed, before supersymmetry breaking, the components of the
St\"uckelberg multiplet form, together with the vector multiplet of
the anomalous gauge boson, a massive vector multiplet of mass
$M_{St}$. Such a multiplet is composed of the massive anomalous gauge boson, whose
mass is given by the St\"uckelberg term, a massive saxion and a
massive Dirac fermion of mass $M_{St}$.  The Dirac fermion is obtained by
diagonalizing the 2-dimensional mass matrix constructed in the basis
of $\lambda_B$ - the gaugino from the vector multiplet $\hat{B}$ - and
$\psi_b$, which is the axino of the St\"uckelberg multiplet. The
diagonalization of this matrix trivially gives two Weyl eigenstates of
the same mass $M_{St}$, which can be assembled into a single massive
Dirac fermion of the same mass.

\subsection{The scalar sector}
The axion and the saxion contribute to the CP-even and CP-odd scalar sectors of the model. One obtains a charged CP-even Higgs sector which involves the states
$(\textrm{Re}H_2^{1},\textrm{Re}H_1^{2})$. The mass matrix has one zero eigenvalue corresponding to a charged Goldstone boson and a mass eigenvalue corresponding to the charged Higgs mass
\begin{align}
m^{2}_{H^\pm}=\left(\frac{v_{1}}{v_{2}}+\frac{v_{2}}{v_{1}}\right)
\left(\frac{1}{4}g^{2}v_{1}v_{2}-\frac{1}{2}\lambda^{2}v_{1}v_{2}
+a_{\lambda}\frac{v_{S}}{\sqrt{2}}\right)
\end{align} 
where $g^2=g_2^2+g_Y^2$ is the combination of the $SU(2)$ and $U(1)_Y$ gauge couplings. In the expression above $v_1$, $v_2$ and $v_s$ are the vacuum expectation values of the of the CP-even scalar component of the two Higgses, $H_1$ and $H_2$ and of the singlet $S$.\\
These two charged states are accompanied by other neutral mass eigenstates derived from the basis $(\textrm{Re}H_1^1,\textrm{Re}H_2^2,\textrm{Re}S,\textrm{Re}b)$.
The four physical states obtained in this sector are
denoted as $H_0^1, H_0^2, H_0^3$ and $H_0^4$. Together with the charged
physical states extracted before, $H^{\pm}$, they describe the 6
degrees of freedom of the CP-even sector.\\
The CP-odd sector is generated
from the basis $(\textrm{Im}H_1^1,\textrm{Im}H_2^2,\textrm{Im}S,
\textrm{Im}b)$.  One obtains two physical states, $H_0^4$ and $H_0^5$,
and two Goldstone modes that provide the longitudinal degrees of
freedom for the neutral gauge bosons, $Z$ and $Z^{\prime}$.

\subsection{Chargino sector}
\label{sec:chargino} 
The chargino sector of the model can also be investigated in detail.  One defines
\begin{equation} \lambda_{w^+}=\frac{1}{\sqrt{2}}(\lambda_{w_1} - i
\lambda_{w_2})\hspace{1cm}\lambda_{w^-}=\frac{1}{\sqrt{2}}(\lambda_{w_1}
+ i \lambda_{w_2}),
\end{equation} 
parameterized by the gaugino mass term $M_W$ and the scalar vevs $v_1,v_2$ and $v_S$. In the basis
$\left(\lambda_{w^+},\tilde{H}_{2}^{1},\lambda_{w^-},\tilde{H}_{1}^{2}\right)$ one derives the mass matrix
\begin{equation} M_{\tilde{\chi}^{\pm}}^{2}=
\begin{pmatrix} 0 & 0 & M_W & g_2 v_1\\ 0 & 0 & g_2 v_2 & \lambda
v_S\\ M_W & g_2 v_2 & 0 & 0\\ g_2 v_1 & \lambda v_S & 0 & 0
\end{pmatrix}
\end{equation} 
which, after diagonalization generates the eigenvalues
\begin{equation} m_{\tilde{\chi}^{\pm}_{1,2}}=\frac{1}{2}\left[M_W^{2}
+ \lambda^{2}v_S^2 + g_2^2 v^2 \mp \sqrt{\left(M_W^{2} +
\lambda^{2}v_S^2 + g_2^2 v^2\right)^{2} - 4\left(\lambda v_S M_W
-g_2^2 v_1 v_2\right)^2}\right].
\end{equation} 
The two mass eigenstates can be defined as
\begin{equation} \chi^{+}=V\psi^{+}\hspace{1cm}\chi^{-}=U\psi^{-}
\end{equation} where $U$ and $V$ are two unitary matrices that perform
the diagonalization from the interaction eigenstates
 
\begin{equation} \psi^{+}= \left(
\begin{array}{c} \lambda_{w^+}\\ \tilde{H}_{2}^{1}
\end{array} \right)\hspace{1cm} \psi^{-}= \left(
\begin{array}{c} \lambda_{w^-}\\ \tilde{H}_{1}^{2}
\end{array} \right).
\end{equation} 
The two matrices $U$ and $V$ satisgfy the relations
\begin{equation} V X^\dagger X V^{-1}=U^* X X^\dagger U^T =
M_{\chi^{\pm}, diag};
\end{equation} where $M_{\chi^{\pm}, diag}$ is given by
\begin{equation} M_{\chi^{\pm}, diag}= \left(
\begin{array}{cc} m_{\tilde{\chi}^{\pm}_{1}} & 0\\ 0 &
m_{\tilde{\chi}^{\pm}_{2}}
\end{array} \right),
\end{equation}
with 
\begin{equation} X=
\begin{pmatrix} M_W & g_2 v_2\\ g_2 v_1 & \lambda v_S
\end{pmatrix}.
\end{equation}

\subsection{Neutralino mass matrix}
\label{sec:neutralino} 
Turning to the neutralino sector, the mass matrix is expressed in the basis
$(i\lambda_{w_3},i\lambda_{Y},i\lambda_{B},\tilde{H}_1^1,\tilde{H}_2^2,\tilde{S},\psi_{\bf
b})$ corresponding to the neutral gauginos $\lambda$'s, which are the fermions of the 
$W_3$, hypercharge and anomalous gauge boson $B$, respectively, accompanied by the two higgsinos $\tilde{H}_i^j$, the singlino $\tilde{S}$ and the axino $\psi_{\bf b}$. The structure of the matrix is summarised below

\begin{equation}
\label{mmchi}
 M_{\chi^0}=
\begin{pmatrix} M_{\chi^0}^{\,\,11} & 0 & 0 & M_{\chi^0}^{\,\,14} &
M_{\chi^0}^{\,\,15} & 0 & M_{\chi^0}^{\,\,17}\\ \cdot &
M_{\chi^0}^{\,\,22} & M_{\chi^0}^{\,\,23} & M_{\chi^0}^{\,\,24} &
M_{\chi^0}^{\,\,25} & 0 & 0\\ \cdot & \cdot &
M_{\chi^0}^{\,\,33} & M_{\chi^0}^{\,\,34} & M_{\chi^0}^{\,\,35} &
M_{\chi^0}^{\,\,36} & M_{\chi^0}^{\,\,37}\\ \cdot & \cdot & \cdot & 0
& M_{\chi^0}^{\,\,45} & M_{\chi^0}^{\,\,46} & 0\\ \cdot & \cdot &
\cdot & \cdot & 0 & M_{\chi^0}^{\,\,56} & 0\\ \cdot & \cdot & \cdot &
\cdot & \cdot & 0 & 0\\ \cdot & \cdot & \cdot & \cdot & \cdot & \cdot
& M_{\chi^0}^{\,\,77}\\
\end{pmatrix}
\end{equation} 
with 

\begin{align} 
M_{\chi^0}^{\,\,11} &= M_{w} \hspace{1cm}
M_{\chi^0}^{\,\,14} = -\frac{g_2 v_1}{2} \hspace{1cm}
M_{\chi^0}^{\,\,15} = \frac{g_2 v_2}{2} \hspace{1cm}
M_{\chi^0}^{\,\,17} = 0
\nonumber\displaybreak[0]\\ 
M_{\chi^0}^{\,\,22} &=M_{Y} 
\hspace{1cm} 
M_{\chi^0}^{\,\,23} = \frac{1}{2} M_{YB}
\hspace{1cm} 
M_{\chi^0}^{\,\,24} = \frac{g_Y v_1}{2}
\hspace{1cm} 
M_{\chi^0}^{\,\,25} = -\frac{g_Y v_2}{4} 
\nonumber\displaybreak[0]\\ 
M_{\chi^0}^{\,\,33} &=\frac{1}{2} M_B
\hspace{1cm} 
M_{\chi^0}^{\,\,34} = -v_1 g_B B_{H_1}  \hspace{1cm} 
M_{\chi^0}^{\,\,35} = -v_2 g_B B_{H_2} \hspace{1cm} 
M_{\chi^0}^{\,\,36} = -v_S g_B B_S  \nonumber\displaybreak[0]\\ 
M_{\chi^0}^{\,\,37} &= M_{St} \hspace{1cm} 
M_{\chi^0}^{\,\,45} = \frac{\lambda v_S}{\sqrt{2}} \hspace{1cm} 
M_{\chi^0}^{\,\,46} = \frac{\lambda v_2}{\sqrt{2}} \hspace{1cm} 
M_{\chi^0}^{\,\,56} = \frac{\lambda v_1}{\sqrt{2}} \hspace{1cm} 
M_{\chi^0}^{\,\,77} = - M_{b}
\end{align} 

and can be investigated numericallly quite effectively. We have denoted with
$M_{\chi_0}$ the corresponding mass matrix. The neutralino eigenstates of this mass matrix are
labelled as $\tilde\chi_i^0$ ($i=0,\dots,6$) and can be expressed in the same basis as
\begin{align}
\tilde{\chi}^{0}_i=a_{i1}\,i\lambda_{W_3}+ a_{i2}\,i\lambda_{Y}+
a_{i3}\,i\lambda_{B} + a_{i4}\,\tilde{H}^1_1 + a_{i5}\,\tilde{H}^2_2 +
a_{i6}\,\tilde{S} + a_{i7}\,\psi_{\bf b}.  
\end{align} 

The neutralino mass eigenstates are ordered in mass and the lightest
eigenstate corresponds to $i=0$. 
The rotation matrix that diagonalizes the
neutralino mass matrix is denoted as $O^{\,{\chi}^0} $, implicitly defined
as
\begin{equation}
\begin{pmatrix} 
i\lambda_{w_3}\\ 
i\lambda_{Y}\\ 
i\lambda_{B}\\
\tilde{H}_1^1\\ 
\tilde{H}_2^2 \\
\tilde{S}\\ 
\psi_{\bf b}
\end{pmatrix}= 
O^{\chi^0}
\begin{pmatrix} 
\chi^0_0\\ 
\chi^0_1\\ 
\chi^0_2\\ 
\chi^0_3 \\ 
\chi^0_4\\ 
\chi^0_5 \\ 
\chi^0_6
\end{pmatrix}.
\end{equation}
Originally, the analysis of the spectrum of this model was performed in the TeV region, having been formulated as a possible extension of the Standard Model with a low scale for supersymmetry. Even if this study is valid in a certain region of parameter space, the main steps followed are fully general and can be applied in other similar formulations as well, possibly with a larger content of scalar superfields and a more general superpotential. It is clear, however, that in such models dark matter with a supersymmetric St\"uckelberg multiplet generates two dark matter candidates, an axion and a neutralino. Notice that the lowest neutralino can have a significant axino component, coming from the same supermultiplet. Detailed simulations of the relkic densities of dark matter for this model have been presented in \cite{Coriano:2010ws}.

\subsection{Parameter choice in the original USSM-A (TeV) model}
To be specific, here, as an example, we consider the case of a St\"uckelberg mass in the TeV region, which illustrates how it is possible to characterize the parameter space of the model, which may be a template for further extensions. We need to fix some of the parameters, first of all by requiring their compatibility with the typical masses of the Standard
Model particles. In the case of $M_{St}$ in the TeV region, the Higgs vev's $v_1$ and $v_2$
have been constrained in order to generate the correct mass values of
the $W^{\pm}$, which depends on $v^2=v_1^2+v_2^2$, and of the $Z$
gauge boson.\\
The choice of $v_S$ and of the parameter $\lambda$
in the trilinear term $\lambda S H_1\cdot H_2$ in the scalar potential
had been made in order be in agreement with the Standard Model Higgs (with $\lambda \sim
O(1)$). Typical values are
$\lambda=0.5$ and the assignment $B_{H_1} = -1, B_{S} = 3$ for
the $U(1)_B$ charges of the Higgs and the singlet; $B_{Q} = 2$ for the  quark doublet, and $B_{L} = 1$ for the lepton doublet. The gauge mass terms had been selected according to the relation
\begin{align} 
M_Y:M_W:M_G=1:2:6,
\label{scelta0} 
\end{align}
coming from the unification
condition for the gaugino masses. As a further simplification, the sfermion mass parameters
$M_L$, $M_Q$, $m_R$, $m_D$ and $m_U$ have been set to a unique value
$M_0$. We have also chosen a common value $ a_0$ for the trilinear
couplings $a_e$,$a_d$ and $a_u$. With these choices, other free
parameters left are the St\"uckelberg mass $M_{St}$, the gaugino mass
term for $\lambda_B$, denoted by $M_B$, and the axino mass term,
$M_b$. Our choices are the following
\begin{align} 
&M_Y=500\textrm{GeV} \hspace{.5cm}
M_{YB}=1\, \textrm{TeV}\hspace{.5cm} 
M_W=1 \,\textrm{TeV} \hspace{.5cm} 
M_G=3\, \textrm{TeV}\nonumber\\
&M_L=M_Q=m_R=m_D=m_U=M_0=1\,\textrm{TeV}\nonumber\\
&a_e=a_d=a_u=a_0=1\,\textrm{TeV}\nonumber\\
&a_\lambda=-100\,\textrm{GeV},
\label{scelta1}
\end{align} 
where with $M_L$ and $M_Q$ we have denoted the scalar mass
terms for the sleptons and the squarks, assumed to be equal for all the 3 generations.  
We have also chosen 
\begin{align} 
M_B=M_b=1\,\textrm{TeV}
\label{terza} 
\end{align}
with a coupling constant $g_B$ of the anomalous
$U(1)$ of $0.4$. 
We also assume a value $v_b=20$ GeV for the vev of the saxion field Re$b$. The most significant parameters in the relic density calculation are $M_{St}$ and the Higgs vev ratio $\tan\beta$.
Concerning the St\"uckelberg mass, in such original analysis, its value had been chosen in two  different regions, $2-10$ TeV and $11-25$ TeV. In both regions one can consider different values of $\tan\beta$, which controls the relation between the two vevs of the scalar sector $v_1, v_2$. With this parameter choice one can extract the decay rate of a 
supersymmetric axion.

\begin{figure}[t]
\begin{center}
\includegraphics[scale=1]{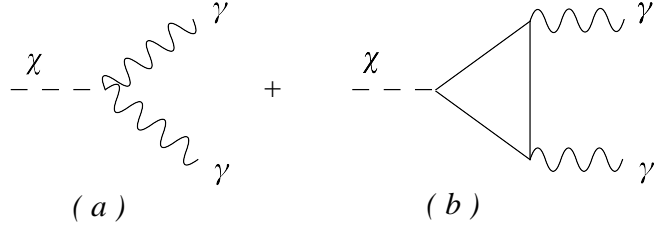}
\caption{ Contributions to the $\chi \rightarrow \g \g$ decay,\label{fig:chi_decay} corresponding to the anomaly contribution (a) and to the interaction of the axion with the fermions, mediated by the Yukava couplings (b). }
\end{center}
\end{figure}
\begin{figure}[t]
\begin{center}
\includegraphics[scale=0.3, angle=-90]{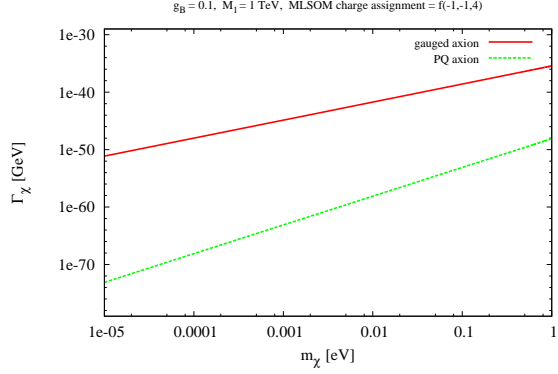}
\end{center}
\caption{\small Total decay rate of the axi-Higgs for several mass values.
Here, for the PQ axion, we have chosen $f_a=10^{10}$ GeV. $g_B$ defines the coupling of the anomalous gauge boson $B$.The values $f(-1,-1,4)$ characterize the charge assignments of the model \cite{Coriano:2010ws}. }
\label{decs1}
\end{figure}

\section{Decay of a gauged supersymmetric axion}
\label{sec:decays-supersymm-gau}  The decay
rate of the axion into two-photons in this supersymmetric model is mediated both by the direct PQ interaction and by the fermion loop,
which are shown in Fig.~\ref{fig:chi_decay}, keeping the axion mass as a free parameter.  
Denoting with $N_c(f)$ the color factor for a fermion flavour ($f$), and introducing the function $\tau_f \,\eta(\tau_f)$, related to the mass of the fermions circulating in the loop with
\begin{equation} \tau=4m_f^2/m_{\chi}^2\hspace{1cm}
\eta(\tau)=\arctan^2 \frac{1}{\sqrt{- \rho_{f
\chi}^2}}\hspace{1cm}\rho_{f \chi} = \sqrt{1 - \left( \frac{2
m_f}{m_\chi} \right)^2 },
\end{equation} the WZ interaction in Fig.~\ref{fig:chi_decay} is given
by \bea {\mathcal M}^{\mu \nu}_{WZ}(\chi \rightarrow \g \g) = 4
g^{\chi}_{\g\g} \varepsilon[\mu,\nu,k_1,k_2], \eea where
$g^{\chi}_{\g\g}$ is the coupling of the physical component of $\textrm{Im} b$ to the photons, defined via the relations
\begin{equation} g^{\chi}_{\g \g}= -\frac{g_B B_S\left(4 c_Y g_2^2 +
c_W g_Y^2\right)}{16 g^2 M_{St}}\sqrt{\frac{v^2 v_S^2+v_1^2 v_2^2}{4
M_{St}^2 v^2 + g_B^2 B_S^2 \left(v^2 v_S^2+v_1^2 v_2^2\right)}}
\end{equation} obtained from the rotation of the WZ vertices on the
physical basis.  The massless contribution to the decay
rate coming from the WZ counterterm $\chi F^{}_\g F^{}_\g$ is given by the relation
\beq
\label{WZrate1} \Gamma^{}_{WZ}(\chi \rightarrow \g\g)=
\frac{m^3_\chi}{4 \pi}(g^\chi_{\g\g})^2.  \eeq
 Combining also in this
case the tree-level decay with the 1-loop amplitude, we obtain for
$\chi \rightarrow \gamma\gamma$ the amplitude 
\beq {\mathcal M}^{\mu \nu}(\chi
\rightarrow \g \g) = {\mathcal M}^{\mu \nu}_{WZ}+{\mathcal M}^{\mu
\nu}_{f}.  \eeq

The second amplitude in Fig.~\ref{fig:chi_decay} is mediated by the
triangle loops and is given by the expression
\begin{equation} 
{\mathcal M}^{\mu \nu}_{f}(\chi \rightarrow \g \g) =
\sum_f N_c(f) \,i C_0(m^2_\chi,m_f) c^{\chi, f}_{\g\g}
\varepsilon[\mu,\nu,k_1,k_2] \hspace{.5cm}
f=\{q_u,q_d,\nu_l,l,\chi_1^{\pm},\chi_2^{\pm}\}
\label{pseudo}
\end{equation} 
where $N_c(f)$ is the color factor for the fermions. In
the domain $0< m_\chi < 2 m_f$, which is the relevant domain for our
study, being the axion very light, the pseudoscalar triangle when both
photons are on mass-shell is given by the expression
\begin{equation} 
C_0(m^2_\chi,m_f) = -\frac{m_f}{\pi^2 m_\chi^2}
\arctan^2 \left( \left( \frac{4 m_f^2}{m_\chi^2}
-1\right)^{-1/2}\right) = -\frac{m_f}{\pi^2 m_\chi^2} \eta(\tau)
\label{region1}
\end{equation} 
The coefficient $c^{\chi, f}_{\g\g}$ is the factor for
the vertex between the axi-Higgs and the fermion current. The
expressions of these factors are
\begin{align}
c^{\chi,q_u}&=-\frac{i\,\sqrt{2}\,y_u\,M_{St}\,v_1^2\,v_2}{\sqrt{(v^2
v_S^2+v_1^2v_2^2)\left[4 M_{St}^2 v^2 + g_B B_S (v^2
v_S^2+v_1^2v_2^2)\right]}},\nonumber\\
c^{\chi,q_d}&=-\frac{v_2\,y_d}{v_1\,y_u}\,c^{\chi, q_u},\nonumber\\
c^{\chi,l}&=\frac{y_e}{y_d}\,c^{\chi, q_d}.
\end{align} 
On obatins the following expression for the decay amplitude
\begin{align} 
\Gamma_\chi\equiv\Gamma(\chi \rightarrow \g\g) =&
\frac{m^3_\chi}{32 \pi} \left\{ 8 (g^\chi_{\g\g})^2 + \frac{1}{2}
\left| \sum_f N_c(f) i \frac{\tau_f~\eta(\tau_f)}{4\pi^2 m_f} e^2
Q_f^2 c^{\chi, f} \right|^2 \right.  \nonumber\\ & \left. \qquad\quad
+ 4 g^\chi_{\g\g} \sum_f N_c(f) i \frac{\tau_f~\eta(\tau_f)}{4\pi^2
m_f} e^2 Q_f^2 c^{\chi, f} \right\},
\end{align} where the three terms correspond, respectively, to the
point-like WZ term, the 1-loop contribution and to their
interference.\\
It is important to observe that in the expression of this decay rate both the direct
($\sim {(g^\chi_{\g\g}})^2$) and the interference ($\sim
g^\chi_{\g\g}$) contributions are suppressed as inverse powers of the
St\"uckelberg mass. For instance, one can choose
$v$ as the SM electroweak vev, while $v_S$  $\sim 500\,\textrm{GeV}$. In order to have an acceptable Higgs spectrum,
the Yukawa couplings have been set to give the right fermion masses of
the Standard Model, while for $g_B$ and $B_S$ we have chosen $g_B=0.1$
and $B_S=4$. Figs.~\ref{decs1} shows the decay rate results for the decay rate as a function of the mass of the
axion $m_\chi$, which clearly indicates that they are very
small for a milli-eV particle, although larger than those of the PQ
case \cite{Coriano:2010py}. One may conclude that a PQ-like axion is indeed long-lived also in
these models and as such could, in principle, contribute to the relic densities of dark matter. 


\section{ Relic densities for a decoupling at the weak scale}
\label{relics} 
The study of the relic densities of such particles can be performed using an approach based on simple considerations from kinetic theory using Boltzmann's equation, combined with the usual counting of the degrees of freedom of a certain theory at the various cosmological epochs. We start from the Lagrangian 
\begin{align} 
\mathcal{S}=
\int d^4 x \sqrt{g}\left( \frac{1}{2}\dot{\chi}^2 - \frac{1}{2}
m_\chi^2 \Gamma_\chi \dot{\chi}\right), 
\end{align} 
where $\Gamma_\chi$ is
the decay rate of the axion and we have expanded the potential around
its minimum up to quadratic terms. The same action is derived from the
quadratic approximation to the general expression 
\begin{align}
\mathcal{S}=\int d^4 x R^3(t)\left( \frac{1}{2} \sigma_{\chi}^2
\left(\partial_{\alpha} \theta\right)^2 - \mu^4 \left(1-
\cos\theta\right) - V_0\right) 
\end{align} 
which, in our case, is constructed
from the expression of $V^\prime$ given in Eq.~(\ref{extrap}), with
$\mu\sim v$. In the models discussed above this scale was the electroweak scale. We also set other contributions to
the vacuum potential to vanish ($V_0=0$). In a
Friedmann-Robertson-Walker spacetime metric with a scaling factor
$R(t)$, this action gives the equation of motion 
\begin{align}
\frac{d}{dt}\biggl[\left( R^3(t) (\dot{\chi} +
\Gamma_\chi\right)\biggr] + R^3 m_{\chi}^2(T) =0.
\label{FRWequation}  
\end{align}
We will neglect the decay rate of the axion
in this case and set $\Gamma_\chi\approx 0$. At this point, since the
potential $V^\prime$ is of non-perturbative origin, we can assume that
it vanishes far above the electroweak scale (or temperature
$T_{ew}$). For this reason $m_\chi=m_b=0$ for $T\gg T_{ew}$, which is
essentially equivalent to the requirement that the St\"uckelberg field is not
subject to any mixing far above the weak scale.  The general equation
of motion derived from Eq.~(\ref{FRWequation}), introducing a
temperature dependent mass, can be written as
\begin{align} 
\ddot{\chi} + 3 H \dot{\chi} + m_{\chi}^2(T) \chi =0,
\label{motioneq}
\end{align} 
which clearly allows as a solution a constant value of
the misalignment angle $\theta=\theta_{i}$. The T-dependence of the
mass term should be generated, for consistency, from a generalization
to finite temperature of $V^\prime$. 
The axion energy density is given by 
\begin{align}
\rho=\frac{1}{2}
\dot{\chi}^2 + \frac{1}{2} m_\chi^2 \chi^2,
\label{rhoeq}  
\end{align}
which after a harmonic averaging gives 
\begin{align}
\langle
\rho\rangle = m_\chi^2 \langle \chi^2\rangle.
\label{averageeq}  
\end{align}
Differentiating Eq.~(\ref{rhoeq}) and using 
the equation of motion in~(\ref{motioneq}), followed by the 
averaging Eq.~(\ref{averageeq}) one obtains the relation
\begin{align} 
\langle \dot{\rho}\rangle =\langle \rho\rangle \left(
-3 H + \frac{\dot{m}}{m}\right),
\end{align} 
where the time dependence of the mass is through its
temperature $T(t)$, while $H(t)=\dot{R}(t)/R(t)$ is the Hubble
parameter. By inspection one easily finds that the solution of this
equation is of the form
\begin{align} 
\langle \rho\rangle = \frac{m_\chi(T)}{R^3(t)}
\end{align} 
showing a dilution of the energy density with an
increasing space volume, valid even for a $T$-dependent mass. At this
point, the universe must be (at least) as old as the required period
of oscillation in order for the axion field to start oscillating and
to appear as dark matter, otherwise $\theta$ is misaligned but frozen. This information 
is the content of the condition
\begin{align} 
m_\chi(T_i)= 3 H(T_i),
\label{mhcond}
\end{align} which allows to identify the initial temperature of the
coherent oscillation of the axion field $\chi$, $T_i$, by equating
$m_\chi(T)$ to the Hubble rate, taken as a function of temperature.

To quantify the relic densities at the current temperature $T_0$
($T_0\equiv T(t_0)$, at current time $t_0$) we need to define
the two effective couplings
\begin{align} 
& g_{*,S,T}=\sum_B g_i \left(\frac{T_i}{T}\right)^3 +
\frac{7}{8}\sum_F g_i \left(\frac{T_i}{T}\right)^3 \nonumber \\ 
& g_{*,T}=\sum_B g_i \left(\frac{T_i}{T}\right)^4 + \frac{7}{8}\sum_F g_i \left(\frac{T_i}{T}\right)^4,
\end{align} 
functions of the massless relativistic degrees of freedom
of the primordial state, with $T\gg T_{ew}$. Coming to the counting of the
degrees of freedom, we have a cotribution equal to 2 for a Majorana fermion and for a massless
gauge boson, 3 for a massive gauge boson and 1 for a real scalar. In
the radiation era, the thermodynamics of all the components of the
primordial state is entirely determined by the temperature $T$, being
the system at equilibrium. We exclude for simplicity all possible inhomogeneities  \cite{Lazarides:1990xp}), in the distribution of the entropy 
and assume no heat transfer, with conservation of the comoving entropy. Pressure and entropy are then just
given as a function of the temperature
\begin{align} & \rho=3 p=\frac{\pi^2}{30} g_{*,T}T^4 \nonumber \\ &
s=\frac{2 \pi^2}{45} g_{*,S,T} T^3,
\label{entropy}
\end{align} while the Friedmann equation allows to relate the Hubble
parameter and the energy density
\begin{equation} H=\sqrt{\frac{8}{3} \pi G_N \rho},
\label{hubble}
\end{equation} with $G_N={1}/{M_P^2}$ being the Newton constant and
$M_P$ the Planck mass. The number density of axions $n_\chi$ decreases
as $1/R^3$ with the expansion, as does the entropy density $s\equiv
S/R^3$, where $S$ indicates the comoving entropy density - which
as already mentioned, remains constant in time ($\dot{S}=0$) - leaving the ratio $Y_a\equiv
n_\chi/s$ conserved. We define, as usual, the abundance variable of
$\chi$
\begin{equation} Y_\chi(T_i)= \frac{n_\chi}{s}\bigg\vert_{T_i}
\end{equation} at the temperature of oscillation $T_i$, and observe
that at the beginning of the oscillations the total energy density is
just the potential one
\begin{equation} \rho_\chi=n_\chi(T_i) m_\chi(T_i)=1/2
m_\chi^2(T_i)\chi_i^2.
\end{equation} We then obtain for the initial abundance at $T=T_i$
\begin{equation} Y_\chi(T_i)= \frac{1}{2}\frac{m_\chi(T_i)
\chi_i^2}{s}= \frac{45 m_\chi(T_i)\chi_i^2}{4 \pi^2 g_{*,S,T} T_i^3}
\label{yeq}
\end{equation} where we have inserted at the last stage the expression
of the entropy of the system at the temperature $T_i$ given by
Eq.~(\ref{entropy}).  At this point, plugging the expression of $\rho$
given in Eq.~(\ref{entropy}) into the expression of the Hubble rate as
a function of density given n Eq.~(\ref{hubble}), the condition for
oscillation Eq.~(\ref{mhcond}) allows to express the axion mass at
$T=T_i$ in terms of the effective massless degrees of freedom
evaluated at the same temperature, that is
\begin{equation} m_\chi(T_i)=\sqrt{\frac{4}{5}\pi^3
g_{*,T_i}}\frac{T_i^2}{M_P}.
\label{Tmass}
\end{equation} This gives for Eq.~(\ref{yeq}) the expression
\begin{equation} Y_\chi(T_i)= \frac{45
\sigma_\chi^2\theta_i^2}{2\sqrt{5 \pi g_{*, T_i}} T_i M_P},
\label{ychi}
\end{equation} where we have expressed $\chi$ in terms of the angle of
misalignment $\theta_i$ at the temperature when oscillations start. We
assume that $\theta_i=\langle \theta\rangle$ is the zero mode of the
initial angle of misalignment after an averaging.  As we have already
mentioned, $T_i$ should be determined consistently by
Eq.~(\ref{mhcond}). However, the presence of two significant and
unknown variables in the expression of $m_\chi$, which are the
coupling of the anomalous $U(1)$, $g_B$, and the St\"uckelberg mass
$M$, forces us to consider the analysis of the T-dependence of $\chi$
phenomenologically less relevant. It is more so if the St\"uckelberg
mass is somehow close to the TeV region, in which case the zero
temperature axion mass $m_\chi$ acquires corrections proportional to
the bare coupling $\left( m_\chi\sim \lambda v(1 + O(g_B)\right)$.

For this reason, assuming that the oscillation temperature $T_i$ is
close to the electroweak temperature $T_{ew}$, Eq.~(\ref{Tmass})
provides an upper bound for the mass of the axion at which the
oscillations occur, assuming that they start around the electroweak
phase transition. In other words, mass values of $\chi$ such that
$m(T_i)\ll 3 H(T_i)$ correspond to frozen degrees of freedom of the
axion at the electroweak scale. This approximation allows to define the oscillation mass in terms of the Hubble
parameter for each given temperature. The relic density due to misalignment can be extracted
from the relations
\begin{equation} \Omega_\chi^{mis}\equiv \frac{\rho_{\chi
0}^{mis}}{\rho_c} =\frac{(n_{\chi 0}
m_\chi)}{\rho_c}=\left(\frac{n_{\chi\,0}}{s_0}\right)\frac{m_\chi
s_0}{\rho_c}
\label{resid}
\end{equation} where we have denoted with $n_{\chi 0}$ the current
number density of axions and with $\rho^{mis}_{\chi 0}$ their current
energy density due to vacuum misalignment. This expression can be
rewritten as
\begin{equation} \Omega_\chi^{mis}=\frac{n_\chi}{s}\bigg\vert_{T_i}
m_\chi\frac{s_0}{\rho_c}
\label{omegaeq}
\end{equation} using the conservation of the abundance $Y_{a
0}=Y_{a}(T_i)$. In Eq.~(\ref{omegaeq}) we have neglected a
possible dilution factor $\gamma=s_{osc}/s_0$ which may be present due
to entropy release. In the expression above, we have introduced the variable
\begin{equation} \rho_c= \frac{3 H_0^2}{8 \pi G_N}
\end{equation} which is the critical density and
\begin{equation} s_0= \frac{2 \pi^2}{45} g_{* S, T_0} T_0^3
\end{equation} which is the current entropy density. In order to determine $g_{* S,
T_0}$, we just recall that at the current temperature $T_0$ the
relativistic species contributing to the entropy density $s_0$ are the
photons and three families of neutrinos with
\begin{equation} g_{* S, T_0}= 2 + \frac{7}{8} \times 3\times 2
\left(\frac{T_\nu}{T_0}\right)^3
\end{equation} where, from entropy considerations,
${T_\nu}/{T_0}=(4/11)^{1/3}$.
In the model we have 13 gauge bosons corresponding to the gauge group
$SU(3)\times SU(2)\times U_Y(1)\times U_B(1)$, 2 Higgs doublets, 3
generations of leptons and 3 families of quarks. Above the energy of
the electroweak transition we have only massless fields with the
exception of the $U_B(1)$ gauge boson, since this symmetry takes the
St\"uckelberg form above the electroweak scale, giving
$g_{*,T}=110.75$. Below the same scale this number is similarly
computed with $g_{*,T}=91.25$. Other useful parameters are the
critical density and the current entropy
\begin{equation}
\rho_{c}=5.2\cdot10^{-6}\textrm{GeV}/\textrm{cm}^3\hspace{1cm}s_0=2970
\,\,\textrm{cm}^{-3},
\end{equation} with $\theta_i\simeq1$.  It is clear, by intersecting
these numbers into Eq.~(\ref{resid}) that $\Omega_\chi^{mis}$ \beq
\frac{45 \sigma_\chi^2\theta_i^2}{2\sqrt{5 \pi g_{*, T_i}} T_i M_P}
\frac{m_\chi s_0}{\rho_c} \eeq is negligible unless $\sigma_\chi\sim
M_{St}^2/v$ is of the same order of $f_a\sim 10^{12}$ GeV, the
standard PQ constant.  One can show that this choice would correspond to
$\Omega_\chi\sim 0.1$, but the value of $M_{St}$ should be of
$O(10^7)$ GeV. Details of this analysis can be found in \cite{Coriano:2010ws}.
 
\section{Ultralight St\"uckelberg axions} 
By raising the St\"uckelberg mass near the Planck or GUT scales, the St\"uckelberg construction acquires a fundamental meaning since it can be directly related to the cancellation of a gauge anomaly 
generated at the same scale \cite{Coriano:2017ghp}. 
An axion could emerge at a certain very high scale and decouple closely to that scale. More recently, these types of analysis  have therefore found their way in the discussion of models which have a far wider field content and gauge structure compared to those that we have reviewed in the previous sections and that we plan to rediscuss in the future. \\
These considerations define a new context in which to harbour such models. 
Then it is natural to identify a consistent formulation of such particles, which turn out to be ultralight, within an ordinary gauge theory. One possible scenario is the one in which the axion emerges at the Planck scale $M_P$, but it acquires a mass at a scale below, 
which in our case is assumed to be the GUT scale. 
Therefore, here we are going to consider an extension of the setup discussed in previous sections, under the assumption that their dynamics is now controlled by two large scales \cite{Coriano:2017ghp}. The first is the scale at which the completion theory lives - for example the Planck scale -  whose effective action takes a local form with the inclusion of a St\"uckelberg field, while the phase transition at which the mechanism of vacuum misalignment takes place is located below, such as the GUT scale. \\
We will consider an $E_6$ based model, derived from $E_8$ \cite{Gross:1984dd} with an
$E(8) \times  E(8)$ symmetry. After a compactification of six spatial dimensions on a Calabi-Yau manifold \cite{Candelas:1985en} the symmetry is reduced to an $E(6)$ GUT gauge theory. 
Other string theory compactifications predict different GUT gauge structures, such as  $SU(5)$ and $SO(10)$. In this model based on $E_6$, however, one can  realize a scenario where two components of dark matter are present.
Fermions are assigned to the ${\bf 27}$ representation of $E(6)$, which is anomaly-free.
Notice that in $E(6)$ a PQ symmetry is naturally present, as shown in \cite{Frampton:1981ik}, which allows to have an ordinary PQ axion, while at the same time it is a realistic GUT symmetry which can break to the SM. This is the gauge structure to which one may append an anomalous $U(1)_X$ symmetry.\\
We will consider a gauge symmetry of the form $E_6\times U(1)_X$, where the gauge boson $B^\mu$ is in the St\"uckelberg phase. $B_{\alpha}$ is the gauge field of $U(1)_X$
and $B_{\alpha\beta} \equiv \partial_{\alpha}B_{\beta} - \partial_{\beta}B_{\alpha}$  the corresponding field strength, while $g_B$ its gauge coupling. As already mentioned, the $U(1)_X$ carries an anomalous coupling to the fermion spectrum. \\
The one-particle irreducible (1PI) effective Lagrangian of the theory at 1-loop level takes a form similar to the one presented in the former models 
\begin{equation}
{\cal L}={\cal L}_{E_6\times U(1)_X}+ {\cal L}_{St} + {\cal L}_{anom} + {\cal L}_{WZ},
\end{equation}
in terms of the gauge contribution of $E_6$ (${\cal L}_{E_6}$), the St\"uckelberg term ${\cal L}_{St}$, the Lagrangian generated at one-loop by the 3-point functions  ${\cal L}_{anom}$, due to the presence of anomalous fermion couplings to the $U(1)_X$ gauge boson, and the Wess-Zumino counterterm (WZ) ${\cal L}_{WZ}$. $\mathcal{L}_{St}$ has been defined as in \eqref{deff}. We have introduced the $E_6$ gauge Lagrangian
\begin{equation}
{\cal L}_{E_6\times U(1)_X}=-\frac{1}{4} F^{(E_6)\, \mu\nu}F^{(E_6)}_ {\mu\nu} -\frac{1}{4} B_{\mu\nu}{B}^{\mu\nu}.
\end{equation}
The WZ contribution is the combination of two terms 
\begin{equation}
{\cal L}_{WZ}= c_1 \frac{b}{M_{St}}F^{(E_6)\, \mu\nu}F^{(E_6)\, \rho\sigma}\epsilon_{\mu\nu\rho\sigma} + 
c_2  \frac{b}{M_{St}}B_{\mu\nu}B_{\rho\sigma}\epsilon^{\mu\nu\rho\sigma} 
\label{WZ}
\end{equation}
needed for the cancellation of the $U(1)_X E_6 E_6$  and $U(1)_X^3$ anomalies, for appropriate values of the numerical constants $c_1$ and $c_2$, fixed by the charge assignments of the model.
In this final form, $M_{St}$ is the mass of the gauge $U(1)_X$ gauge boson, which we can be taken of the order of the Planck scale, guaranteeing the decoupling of the axion around the GUT scale $M_{GUT}$, due to the gravitational suppression of the WZ counterterms.  

The three chiral familes will be assigned under $E(6)\times U(1)_X$ to the representations
\begin{equation}
{\bf 27}_{X_1}  ~~~~~ {\bf 27}_{X_2} ~~~~~ {\bf 27}_{X_3},
\label{chiral}
\end{equation}
in which the charges $X_i$ ($i=1,2,3$) are free parameters, later constrained by the cancellation of 
the $U(1)_X^3$ and $E_6\times U(1)_X^2$ anomalies 
\begin{equation}
\sum_{i=1}^{3} X_i^3 = 0,     \qquad\qquad \sum_{i=1}^3 X_i=0.
\label{Xcharges}
\end{equation}
These need to be violated in order to compensate with a Wess-Zumino term for the restoration of the gauge symmetry of the action.\\
The scalar sector contains two ${\bf 351}_{X_i}$ $(i=1,2)$ irreducible representations, where the $U(1)_X$ charges $X_i$ need to be determined. 
The ${\bf 351}$ is the {\it antisymmetric} part of the Kronecker product ${\bf 27} \otimes {\bf 27}$ where ${\bf 27}$ is the defining representation of $E(6)$. The ${\bf 351}_X$ can be conveniently described by the 2-form $A_{\mu\nu} = - A_{\nu\mu}$ with $\mu,\nu = 1$ to $27$. We write down the 
most general renormalizable potential in ${\cal L}_{E_6}$, expressed in terms of $A^{(1)}_{\mu\nu}$ and $A^{(2)}_{\mu\nu}$ of $U(1)_X$ of charges $x_1$ and $x_2$ respectively. 
If we denote the ${\bf 27}_{X_i}$ of Eq.(\ref{chiral}) by $\Psi_{\mu}$ with $\mu = 1,\ldots$  $27$ then the full Lagrangian including the potential $V$, has an invariance under the global symmetry 
 \begin{equation}
 A^{(1)}_{\mu\nu} \rightarrow e^{i\theta} A^{(1)}_{\mu\nu}~~~~~A^{(2)}_{\mu\nu} \rightarrow e^{i\theta} A^{(2)}_{\mu\nu} ~~~~ \Psi_{\mu} \rightarrow e^{- (\frac{1}{2}i\theta)} \Psi_{\mu}.
 \label{PQ}
 \end{equation}
This is identifiable as a Peccei-Quinn symmetry which is broken at the GUT scale when $E(6)$
 is broken to $SU(5)$ \cite{Frampton:1981ik}. This axionic symmetry can be held responsible for solving the strong CP problem. 
  \noindent
In this model we couple $A^{(1)}_{\mu\nu}$ to the fermion families $({\bf 27})_{X_i}$ $i=1,2,3$. We choose
 in Eq. (\ref{chiral}), {\it e.g.} $X_1=X_2=X_3=+1$, with the $X$-charge of $A^{(1)}$ fixed to $X=-2$. The second scalar representation $A^{(2)}$ is decoupled from the fermions, with an 
 $X-$charge for $A^{(2)}$ which is arbitrary and taken for simplicity to be $X=+2$. 
 The potential is expressed in terms of three $E_6\times U(1)_X$ invariant components, 
\begin{equation}
V= V_1+  V_2 + V_p,
\end{equation}
where
\begin{equation} 
V_1= F(A^{(1)},A^{(1)})  \qquad  V_2= F(A^{(2)},A^{(2)}),
\end{equation}
with $V_1$ and $V_2$ denoting the contributions of $({\bf 351})_{-2}$ and $({\bf 351})_{+2}$, expressed in terms of 
the function \cite{Frampton:1981ik}
\begin{eqnarray}
 F(A^{(i)},A^{(j)}) & = &\left. M_{GUT}^2 A^{(i)}_{\mu\nu} \bar{A^{(j)}}^{\mu\nu} +h_1 ~(A^{(i)}_{\mu\nu} \bar{A^{(j)}}^{\mu\nu})^2  +h_2  ~ A^{(i)}_{\mu\nu} \bar{A}^{\nu\sigma} A^{(i)}_{\sigma\tau}\bar{A}^{\tau\mu} \right.\nonumber \\
&&\left.\qquad \qquad +\,h_3 ~  d^{\mu\nu \lambda} d_{\xi\eta\lambda} A^{(i)}_{\mu\sigma}A^{(i)}_{\nu\tau} \bar{A^{(j)}}^{\xi\sigma} \bar{A^{(j)}}^{\eta\tau} \right. \nonumber \\
 & &\left.  \qquad \qquad +\, h_4 ~ d^{\mu\nu\alpha}d^{\sigma\tau\beta}d_{\xi\eta\alpha} d_{\lambda\rho\beta} A^{(i)}_{\mu\sigma}A^{(i)}_{\nu\tau} \bar{A^{(j)}}^{\xi\lambda} \bar{A^{(j)}}^{\eta\rho}\right. \nonumber \\
 & &\left. \qquad \qquad + \,h_5 ~ d^{\mu\nu\alpha} d^{\sigma\beta\gamma} d_{\xi\eta\beta} d_{\lambda\alpha\gamma} A^{(i)}_{\mu\sigma}A^{(i)}_{\nu\tau} \bar{A^{(j)}}^{\xi\lambda} \bar{A^{(j)}}^{\eta\tau} \right.\nonumber \\
 & &\left. \qquad \qquad +\,h_6 ~ d^{\mu\nu\alpha}d^{\sigma\tau\beta} d_{\alpha\beta\gamma} d^{\gamma\zeta\xi} d_{\xi\eta\zeta} d_{\lambda\rho\chi} A^{(i)}_{\mu\sigma}\bar{A^{(j)}}^{\xi\lambda} A^{(i)}_{\nu\tau} \bar{A^{(j)}}^{\eta\rho}\right. ,
 \end{eqnarray}
 in which $d_{\alpha\beta\gamma}$ with $\alpha,\beta,\gamma = 1$ to $27$ is the $E(6)$
invariant tensor.\\
We can follow a procedure which is quite similar to the one implemented in the case of the  two Higgs doublet model discussed in the previous sections. Also in this case we are allowed to introduce a periodic potential on the basis of the underlying gauge symmetry, which takes the form
\begin{eqnarray}
V_p & = & M_{GUT}^2 A^{(1)}_{\mu\nu} \bar{A^{(2)}}^{\mu\nu}e^{- i 4\frac{b}{M_S}}   + e^{- i 8 \frac{b}{M_S}}\left[(h_1 ~(A^{(1)}_{\mu\nu} \bar{A^{(2)}}^{\mu\nu} )^2  + h_2  ~ A^{(1)}_{\mu\nu} \bar{A^{(2)}}^{\nu\sigma} A^{(1)}_{\sigma\tau}\bar{A^{(2)}}^{\tau\mu} \right.\nonumber \\
&&\left.\qquad \qquad +\,h_3 ~  d^{\mu\nu \lambda} d_{\xi\eta\lambda} A^{(1)}_{\mu\sigma}A^{(1)}_{\nu\tau} \bar{A^{(2)}}^{\xi\sigma} \bar{A^{(2)}}^{\eta\tau} \right. \nonumber \\
 & &\left.  \qquad \qquad +\, h_4 ~ d^{\mu\nu\alpha}d^{\sigma\tau\beta}d_{\xi\eta\alpha} d_{\lambda\rho\beta} A^{(1)}_{\mu\sigma}A^{(1)}_{\nu\tau} \bar{A^{(2)}}^{\xi\lambda} \bar{A^{(2)}}^{\eta\rho}\right. \nonumber \\
 & &\left. \qquad \qquad + \,h_5 ~ d^{\mu\nu\alpha} d^{\sigma\beta\gamma} d_{\xi\eta\beta} d_{\lambda\alpha\gamma} A^{(1)}_{\mu\sigma}A^{(1)}_{\nu\tau} \bar{A^{(2)}}^{\xi\lambda} \bar{A^{(2)}}^{\eta\tau} \right.\nonumber \\
 & &\left. \qquad \qquad +\,h_6 ~ d^{\mu\nu\alpha}d^{\sigma\tau\beta} d_{\alpha\beta\gamma} d^{\gamma\zeta\xi} d_{\xi\eta\zeta} d_{\lambda\rho\chi} A^{(1)}_{\mu\sigma}\bar{A^{(2)}}^{\xi\lambda} A^{(1)}_{\nu\tau} \bar{A^{(2)}}^{\eta\rho}\right] + h.c.
 \end{eqnarray}
and which becomes periodic at the GUT scale after symmetry breaking, similarly to the case considered in 
\cite{Coriano:2010py,Coriano:2010ws}. As already mentioned in the previous sections, this potential is expected to be present at the GUT phase transition. Also in this case the size of the 
 contributions in $V_p$, generated by instanton effects at the GUT scale, are expected to be exponentially suppressed. However, the size of the suppression is related to the value of the gauge coupling at the corresponding scale. This is crucial in order to identify an ultralight axion. 

 \subsection{The periodic potential}
 
The $E_6\times U(1)_X$ symmetry at $M_{GUT}$ can be broken in different ways, such as  
  $E(6) \supset SU(3)_C \times SU(3)_L \times SU(3)_H$ where 
\begin{eqnarray}
\label{351prime}
({\bf 351}) & = & (1, 3^*, 3) + (1, 3^*,6^*) + (1, 6, 3) + (3, 3, 1) + (3, 6^*, 1) + (3, 3, 8) + \nonumber \\
& & (3^*, 1, 3^*) + (3^*, 1, 6)
+ (3^*, 8, 3^*) + (6^*, 3, 1) + (6, 1, 3^*) + (8, 3^*, 3)
\label{351}
\end{eqnarray}
of which the colour singlets are only the 45 states for each of the two $({\bf 351})_{X_i}$
\begin{equation}
(1, 3^*, 3)_{X_i} \qquad   (1, 3^*, 6^*)_{X_i} \qquad (1, 6, 3)_{X_I}  ~~~  i=1,2.
\label{colorsinglets}
\end{equation}
One easily realizes that there are exactly nine colour-singlet $SU(2)_L$-doublets in the $({\bf 351}^{'})_{-2}$ and 9 in the 
$({\bf 351}^{'})_{+2}$, that we may denote as  $H^{(1)}_j$, $H^{(2)}_j$, with $j=1,2\ldots 9$, which appear in the periodic potential in the form 
\begin{eqnarray}
V_p &\sim&\sum_{j=1}^{12}\lambda_0 M_{\textrm {GUT}}^2 (H^{(1)\dagger }_j H^{(2)}_j e^{- 4 i g_B\frac{b}{M_S}})+
\sum_{j,k=1}^{12}\left[\lambda_1(H^{(1)\dagger}_j H^{(2)}_j e^{-i 4 g_B \frac{b}{ M_S}})^2+\lambda_2(H_i^{(1)\dagger}H_i)(H_i^{(1)\dagger}H^{(2)}_j e^{-i 4 g_B\frac{b}{M_S}})\right.\nonumber \\
&&\left. + \lambda_3(H_k^{(2)\dagger}H_k^{(2)})(H_j^{(1)\dagger}H_k^{(2)} e^{-i 4 g_B\frac{b}{M_S}}) \right] +\textrm{h.c.},
\end{eqnarray}
where we are neglecting all the other terms generated from the decomposition (\ref{351prime}) which will not contribute to the breaking. The assumption that such a potential is instanton generated at the GUT scale, 
with parameters $\lambda_i$'s induces a specifc value of the instanton suppression which is drastically different from the case of a St\"uckelberg scale located at TeV/multi TeV range. 

For simplicity we will consider only a typical term in the expression above, involving two neutral components, generically denoted as $H^{(1)\, 0}$ and $H^{(2)\, 0}$, all the remaining contributions being similar. 
In this simplified case the axi-Higgs $\chi$ is generated by the mixing of the CP odd components of two neutral Higgses. The analysis follows rather closely the approach discussed in the previous sections for the simplest two-Higgs doublet model, which defines the template for such constructions.\\
Therefore, generalizing this procedure, the structure of $V_{p}$ after the breaking of the $E_6\times U(1)_X$ symmetry can be summarised in the form 
\begin{align}
V_p\sim&  v_1 v_2
\left(\lambda_2 v_2^2+\lambda_3 v_1^2+\overline{\lambda_0} M_{GUT}^2\right) \cos\left(\frac{\chi}{\sigma_\chi}\right) +  \lambda_1 v_1^2 v_2^2 \cos\left(2\frac{\chi}{\sigma_\chi}\right),
\label{extrap1}
\end{align}
with a mass for the physical axion $\chi$  given by
\begin{equation}
m_{\chi}^2\sim\frac{2 v_1 v_2}{\sigma^2_\chi}\left(\bar{\lambda}_0 v_1^2 +\lambda_2 v_2^2 +\lambda_3 v_1^2+4 \lambda_1 v_1 v_2\right) 
\approx \lambda v^2
\label{axionmass}
\end{equation}
with $v_1\sim v_2\sim v\sim M_{GUT}$. Assuming that $M_S$, the St\"uckelberg 
mass, is of the order of $M_{\textrm{Planck}}$ and that the breaking of the $E_6\times U(1)_X$ symmetry takes place at the GUT scale
$M_{GUT}\sim 10^{15}$ GeV,  (e.g. $v_1\sim v_2\sim M_{GUT}$) then 
\begin{equation}
\sigma_{\chi}\sim M_{\textrm{GUT}} + {\cal O}(M_{\textrm{GUT}}^2/M_{\textrm{Planck}}^2),  \qquad m_\chi^2\sim \lambda_0 M_{\textrm{GUT}}^2,
\end{equation}
where all the $\lambda_i$'s in $V_p$ are of the same order.
The potential $V_p$ being generated by the instanton sector, the size of the numerical coefficients appearing in its expression are constrained to specific values. 
One obtains $\lambda_0\sim e^{-2 \pi/\alpha(M_{\textrm{GUT}})}$, with the value of the coupling  $4 \pi g_B^2 =\alpha_{\textrm{GUT}}$ fixed at the GUT scale. If we assume that  $1/33 \le \alpha_{GUT} \le 1/32$, then  $e^{-201}\sim 10^{-91}\le \lambda_0 \le e^{-205}\sim 10^{-88}$, and
the mass of the axion $\chi$ takes the approximate value 
\begin{equation}
10^{-22}   \textrm{ eV} < m_{\chi} <  10^{-20} \textrm{ eV}, 
\end{equation}
which contains the allowed mass range for an ultralight axion, as discussed in recent analysis of the astrophysical constraints on this type of dark matter \cite{Hui:2016ltb}.\\
\subsection{Conclusions and perspectives} 
We have presented an overview of models based in St\"uckelberg axions, which are formulated in terms of a local effective action - supersymmetric and non - which generalize the PQ construction. The gauge invariance of the effective action is obtained, in the presence of anomalous abelian interactions, by the inclusion of a physical axion-like particle in the spectrum of the theory. As we have illustrated, the crucial idea which takes to a new dark matter candidate is the realization, in a local field theory, of a different mechanism of anomaly cancellation which is not present in the Standard Model or in ordinary anomaly-free gauge theories. If such a cancellation is realized by nature, then dark matter could be the signature of such mechanism. Obviously, this does not exclude that other light or ultralight states, which are ubiquitous in string completions, could not be similarly present. 
However, the constraints that a St\"uckelberg theory imposes on the fermion spectrum are far more specific than those derived from geometric compactifications, which suffer from generic and lack considerable detail at phenomenological level. In these models there is also much less room to play around, but enough 
to move away from the tight constraints which severely limit the parameter space of the ordinary PQ scenario, in all of its variants, where the QCD vacuum defines a strict relation between the axion mass and its coupling to the anomaly.  
\vspace{1cm}

\centerline{\bf Acknowledgements}
We thank Paul Frampton, Marco Guzzi and Antonio Mariano for collaborating to several of these analysis. This work is supported by INFN under Iniziativa Specifica QFT-HEP. 


\end{document}